# Table of Contents

# APML: An Architecture Proof Modeling Language[*]

Diego Marmsoler and Genc Blakqori

Technische Universität München, Germany
diego.marmsoler@tum.de

**Abstract.** To address the increasing size and complexity of modern software systems, compositional verification separates the verification of single components from the verification of their composition. In architecture-based verification, the former is done using Model Checking, while the latter is done using interactive theorem proving (ITP). As of today, however, architects are usually not trained in using a full-fledged interactive theorem prover. Thus, to bridge the gap between ITP and the architecture domain, we developed APML: an architecture proof modeling language. APML allows one to sketch proofs about component composition at the level of architecture using notations similar to Message Sequence Charts. With this paper, we introduce APML: We describe the language, show its soundness and completeness for the verification of architecture contracts, and provide an algorithm to map an APML proof to a corresponding proof for the interactive theorem prover Isabelle. Moreover, we describe its implementation in terms of an Eclipse/EMF modeling application, demonstrate it by means of a running example, and evaluate it in terms of a larger case study. Although our results are promising, the case study also reveals some limitations, which lead to new directions for future work.

## 1 Introduction

Software intensive systems are becoming increasingly big and complex, which makes their verification a challenge. To address this challenge, compositional verification techniques separate the verification of single components from the verification of their composition. In architecture-based verification (ABV) [30], for example, verification of such systems is split into two parts: First, suitable contracts are identified for the involved components and their implementation is verified against these contracts. Since a single component is usually of limited complexity, in ABV this step is fully automated using Model Checking [2]. In a second step, component contracts are combined to verify overall system properties. Reasoning about the composition of contracts, however, might be difficult and sometimes requires manual interaction [29]. Thus, in ABV, it is done using interactive theorem provers, such as Isabelle [33].

---

A full-fledged interactive theorem prover, however, can be quite complex and its usage usually requires expertise which is not always available in the architecture context [28]. Thus, in an effort to bridge the gap between interactive theorem proving and the architecture domain, we developed APML: a language to specify proofs for the composition of contracts using abstractions an architect is familiar with. APML comes with a *graphical notation*, similar to Message Sequence Charts [15], to sketch proofs at the architecture level and it is shown to be *sound* and *complete* regarding the verification of architecture contracts. It is *implemented in Eclipse/EMF* [40], where it can be used to model proofs for architecture contracts and synthesize corresponding proofs for Isabelle's structured proof language Isar [42,43].

The aim of this paper is to introduce APML. To this end, we provide the following contributions: (i) We provide a formal description of APML, including a formal semantics for architecture contracts. (ii) We show soundness and completeness of APML for the verification of architecture contracts. (iii) We present an algorithm to map an APML proof to a corresponding proof in Isabelle/Isar. (iv) We describe its implementation in terms of an Eclipse/EMF modeling application. (v) We demonstrate the approach by means of a running example and report on the outcome of a case study in which we applied APML for the verification of a railway control system. Thereby, to the best of our knowledge, *this is the first attempt to synthesize proofs for an interactive theorem prover from an architecture description*.

Our presentation is structured as follows: In section 2, we provide some background to clarify our understanding of architecture in general and specifically our notion of architecture contract. In section 3, we describe our running example, a reliable calculator. In section 4, we introduce APML, demonstrate it by verifying a property for our running example, and present our soundness and completeness results. In section 5, we present our algorithm to map an APML proof to a corresponding proof in Isabelle/Isar and demonstrate it by means of the running example. In section 6, we describe the implementation of APML in terms of an Eclipse/EMF modeling application and in section 7 we describe our effort to evaluate APML by means of a larger case study. In section 8, we discuss related work before we conclude the paper in section 9 with a brief summary and a discussion of future work.

## 2 Background

### 2.1 Basic Mathematical Notations

For a function $f\colon D \to R$, we shall use $f|_{D'}\colon D' \to R$ to denote the restriction of $f$ to domain $D' \subseteq D$. In addition, we shall use partial functions $f\colon D \dashrightarrow R$ for which we denote with $\mathrm{dom}\,(f) \subseteq D$ its domain and with $\mathrm{ran}\,(f) \subseteq R$ its range.

We will also use finite as well as infinite *sequences* of elements. Thereby, we denote with $(E)^*$ the set of all finite sequences over elements of a given set $E$, by $(E)^\infty$ the set of all infinite sequences over $E$, and by $(E)^\omega$ the set of all finite and infinite sequences over $E$. The $n$-th element of a sequence $s$ is denoted with $s(n-1)$ and the first element is $s(0)$. Moreover, we shall use $\#s \in \mathbb{N}_\infty$ to denote the length of $s$. For a sequence $s \in (D \to R)^\infty$ of functions we shall use $s|_{D'}$ to denote the sequence of all restrictions $s(n)|_{D'}$.

## 2.2 Architecture Model

In our model [27,32], components communicate to each other by exchanging *messages* over *ports*. Thus, we assume the existence of set $\mathcal{M}$, containing all messages, and set $\mathcal{P}$, containing all ports, respectively. Moreover, we postulate the existence of a type function

$$\mathcal{T}\colon \mathcal{P} \to \wp(\mathcal{M}) \tag{1}$$

which assigns a set of messages to each port.

Ports are means to exchange messages between a component and its environment. This is achieved through the notion of port valuation. Roughly speaking, a valuation for a set of ports is an assignment of messages to each port.

**Definition 1 (Port Valuation).** *For a set of ports $P \subseteq \mathcal{P}$, we denote with $\overline{P}$ the set of all possible, type-compatible* port valuations, *formally:*

$$\overline{P} \stackrel{def}{=} \left\{ \mu \in (P \to \mathcal{M}) \mid \forall p \in P\colon \mu(p) \in \mathcal{T}(p) \right\}$$

Components communicate through interfaces by receiving messages on its input ports and sending messages through its output ports.

**Definition 2 (Interface).** *An* interface *is a pair $(I, O)$, consisting of* disjoint *sets of input ports $I \subseteq \mathcal{P}$ and output ports $O \subseteq \mathcal{P}$. For an interface $f$, we denote by $in(f)$ the set of input ports, $out(f)$ the set of output ports, and $port(f)$ the set of all ports. A set of interfaces is called* disjoint *iff its interfaces do not share any ports. For such sets of interfaces, we shall use the same notation as introduced for single interfaces, to denote their input, output, and all ports.*

In addition, a component has a behavior which is given in terms of a *non-empty* set of sequences of port valuations over its interface.

In our model, an architecture connects input and output ports of a set of interfaces. Thereby, the types of connected ports must be compatible.

**Definition 3 (Architecture).** *An* architecture *is a pair $(F, N)$, consisting of a* disjoint *set of interfaces $F$ and a connection $N\colon in(F) \dashrightarrow out(F)$, such that*

$$\forall p \in \mathrm{dom}\,(N) : \mathcal{T}(N(p)) \subseteq \mathcal{T}(p) \tag{2}$$

Note that a connection is modeled as a *partial* function from input to output ports, meaning that not every input port of an architecture is connected to a corresponding output port and vice versa. Thus, ports of an architecture can be classified as either *connected* (given by $\mathrm{dom}\,(N) \cup \mathrm{ran}\,(N)$) or *disconnected* (given by $(in(F) \setminus \mathrm{dom}\,(N)) \cup (out(F) \setminus \mathrm{ran}\,(N))$).

## 2.3 Composition

The interface of an architecture with its environment is given by its disconnected ports.

**Definition 4 (Architecture Interface).** *For an architecture $A = (F, N)$, its interface is defined as $_A\otimes = (I, O)$, consisting of input ports $I = in(F) \setminus \mathrm{dom}(N)$ and output ports $O = out(F) \setminus \mathrm{ran}(N)$.*

Note that, since $F$ is required to be disjoint, an architecture's input and output ports are guaranteed to be disjoint, too. Thus, an architecture interface fulfills all the requirements of definition 2 and thus represents a valid interface. Hence, we can use the same notation as introduced for interfaces to access its ports.

We can now define a notion of composition to obtain the behavior of an architecture from the behavior of its components.

**Definition 5 (Architecture Behavior).** *Given an architecture $A = (F, N)$ and a non-empty behavior $\mathcal{B}_f \subseteq \overline{(port(f))}^\infty$ for all of its interfaces $f \in F$. The behavior of the composition is given by a set of traces $_A\otimes \mathcal{B} \subseteq \overline{(port(_A\otimes))}^\infty$, defined as follows:*

$$_A\otimes \mathcal{B} \stackrel{def}{=} \{t|_{port(_A\otimes)} \mid t \in \overline{(port(F))}^\infty \wedge \tag{3}$$

$$\left(\forall f \in F \colon t|_{port(f)} \in \mathcal{B}_f\right) \wedge \tag{4}$$

$$(\forall (i, o) \in N,\ n \in \mathbb{N} \colon t(n)(i) = t(n)(o))\ \} \tag{5}$$

Roughly speaking, the behavior of a composition is defined as all traces over the architecture's interface (Eq. (3)), which respect the behavior of each component (Eq. (4)) and the connections imposed by the architecture (Eq. (5)).

### 2.4 Contracts

In the following, we are considered with the specification of architectures (as they were described in the previous section). To this end, we assume the existence of a set of predicates $\Gamma(P)$ to specify valuations for a set of ports $P \subseteq \mathcal{P}$.

Our notion of contract is inspired by Dwyer's work on specification patterns [18] which is often found in practice [24]. Thereby, contracts have the form: "if $P$ is true then $Q$ happens after $d$ time points".

**Definition 6 (Contracts).** *A* contract *for an interface $(I, O)$ is a triple $(tg, gr, d)$, consisting of a (possibly empty) trigger $tg \in (\Gamma(I) \times \mathbb{N})^*$, a guarantee $gr \in \Gamma(O)$, and a duration $d \in \mathbb{N}$. For every entry $e$ of a trigger, we denote by $state(e)$ its predicate and with $time(e)$ its time point. Moreover, we require the following conditions for a contract:*
- *The time point of the first trigger is $0$: $time(tg(0)) = 0$ (if $\#tg > 0$).*
- *Triggers are ordered by their time points: $\forall j, j' < \#tg \colon j \leq j' \implies time(tg(j)) \leq time(tg(j'))$.*
- *The guarantee is after the last trigger: $d > time(tg(\#tg-1))$ (or $d > 0$ if $tg = \langle\rangle$).*

Moreover, since they are specified over interfaces, contracts can be specified for components as well as for architectures. They are best expressed graphically using a notation similar to Message Sequence Charts [15] (see fig. 4 or fig. 5 for an example).

In the following, we define what it means for a behavior of a component (or architecture) to satisfy a corresponding contract. Thereby, we denote with $\mu \models \gamma$ that a valuation $\mu \in \overline{P}$ satisfies a predicate $\gamma \in \Gamma(P)$.

**Definition 7 (Satisfaction).** *A behavior $\mathcal{B}$ for an interface* satisfies *a contract $k = (tg, gr, d)$ for that interface, written $\mathcal{B} \models k$, whenever for all $t \in \mathcal{B}$, satisfaction of the triggers implies satisfaction of the guarantee:*

$$\forall n \in \mathbb{N} \colon \Big( (\forall j < \#tg \colon t(n + time(tg(j))) \models state(tg(j))) \implies t(n+d) \models gr \Big)$$

Again, the same definition can be applied for component contracts as well as for architecture contracts.

### 2.5 Isabelle

Isabelle [33] is a generic proof assistant which allows mathematical formulæ to be expressed in a formal language and which provides tools for proving those formulas in a logical calculus. The version of Isabelle used for the work presented in this paper is Isabelle/HOL, which includes tools to support the specification of datatypes, inductive definitions, and recursive functions.

Specifications in Isabelle are grouped into so-called theories, which may import other theories. To modularize results, Isabelle supports the development of abstract specifications by means of locales [3]. Figure 1 shows how such a locale usually looks like: It consists of a name, a list of parameters, and a list of assumptions about these parameters. In previous work [29], we show how to map an architecture specification to a corresponding Isabelle locale. Thereby, ports are mapped to corresponding locale parameters and specifications to locale assumptions.

```
locale name =
  fixes parameter₁ :: 'a ⇒ 'a ⇒ bool
    and parameter₂ :: 'a ⇒ 'a ⇒ bool
    ⋮
  assumes "formula₀"
      and "formula₁"
      ⋮
  begin
    ⋮
  end
```

Fig. 1: A typical Isabelle locale.

In Isabelle, proofs can be expressed in a natural way using Isabelle's structured proof language Isar [43]. A typical Isar proof is depicted in fig. 2: It consists of a sequence of proof steps, which are discharged by some proof methods. For example, Is-

```
proof
  assume label₀: "formula₀"
  from label₀ have label₁: "formula₁" by blast
  ⋮
  from label₀, label₁, … show "formulaₙ" by blast
qed
```

Fig. 2: A typical Isabelle/Isar proof.

abelle's classical reasoner *blast* can perform long chains of reasoning steps to prove formulas. Or the simplifier *simp* can reason with and about equations. Moreover, external, first-order provers can be invoked through *sledgehammer*.

## 3 Running Example: A Reliable Adder

As a running example, let us consider a simple system which calculates the sum of two numbers in a redundant way. Its architecture is depicted in fig. 3: It consists of a dispatcher component which receives two numbers as input from its environment and forwards copies of these numbers to two different adder components. The adder components then calculate the sum of the two numbers and communicate their result to a merger component. The merger component compares the two results and forwards the final result to its environment.

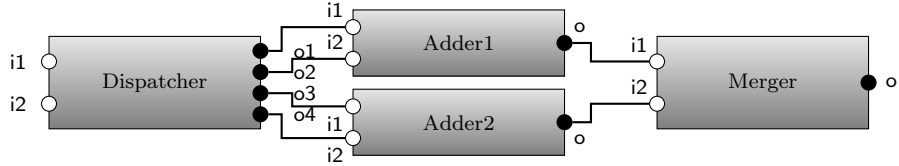

Fig. 3: Architecture for a reliable adder.

The behavior of each component is specified in terms of contracts (as introduced by definition 6) by the sequence diagrams depicted in fig. 4: Figure 4a depicts contract `dispatch` for the dispatcher component. It requires a dispatcher to forward incoming messages received at ports i1 and i2, on its output ports o1 − o4, within one time unit. The contracts for the two adder components, `add1` and `add2`, are depicted in fig. 4b and fig. 4c, respectively. They both require an adder to calculate the sum of the numbers obtained on its input ports i1 and i2 and output it on its output port o. For our example, we assume that the two components use different algorithms to calculate the sum, which is why Adder1 requires four time units while Adder2 requires only three time units to output its result. Figure 4d, fig. 4e, and fig. 4f, depict three different contracts for the merger component. Contract `merge1` requires the merger component to compare the messages received on its input ports i1 and i2, and for the case they coincide, to forward the message after two time units on its output port o. Contracts `merge2` and `merge3` require the merger component to cope with a potential delay of one time unit for messages received on its input ports i2 and i1, respectively.

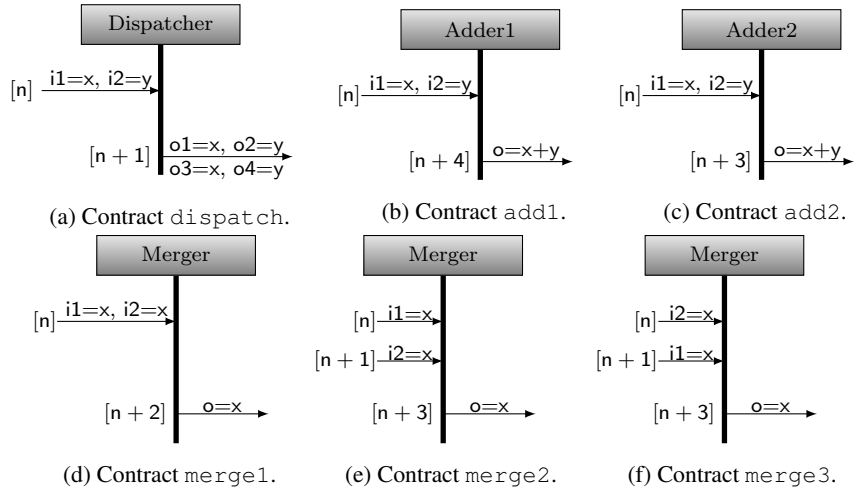

Fig. 4: Contracts for components of reliable adder.

Among other things, we expect the resulting system to output the sum of two numbers it receives on its input ports i1 and i2 after seven time units on its output port o . This can be expressed in terms of a contract over its architecture as specified by the sequence diagram depicted in fig. 5.

Note that our running example is deliberately oversimplified, since its main purpose is to demonstrate our concepts and ideas rather than evaluating the approach in a real world scenario. For details about how the approach works on a real example, we refer to the description of the case study in section 7.

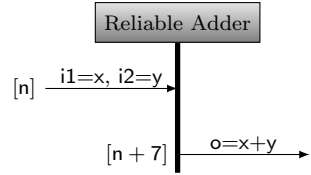

Fig. 5: Contract for reliable adder.

## 4 Modeling Architecture Proofs

An architecture contract can be verified from a set of contracts for its components through a sequence of proof steps.

**Definition 8 (Proof Step).** *A* proof step *for an architecture $(F, N)$ and corresponding contracts $K_f$ for its interfaces $f \in F$, is a 4-tuple $s = (tp, \gamma, r, rf)$, consisting of*
- *a time point $tp \in \mathbb{N}$ denoted $time(s)$,*
- *an architecture state $\gamma \in \Gamma(out(f))$ (for some $f \in F$) denoted $state(s)$,*
- *a rationale $r = (tg, gr, d) \in K_f$ (for some $f \in F$) denoted $rat(s)$,*
- *a (possibly empty) sequence of references $rf \in (R)^*$ denoted $ref(s)$, such that $\#rf = \#tg$ and where $R$ is a* non-empty *set of elements $\mathbb{N} \cup (\mathbb{N} \times \wp(N))$.*

Note that an element of $R$ is either a reference to an assumption $\mathbb{N}$ of the architecture contract we want to prove, or a reference to another proof step $\mathbb{N}$ and a set of connections $\wp(N)$. An architecture proof is given by a sequence of corresponding proof steps.

**Definition 9 (Architecture Proof).** *Given an architecture $A = (F, N)$ and corresponding contracts $K_f$ for its interfaces $f \in F$. An* architecture proof *for a contract $(tg, gr, d)$ over the architecture's interface ${}_A\!\otimes$ is a* finite, non-empty *sequence ps of proof steps, such that the state of the last entry implies the guarantee of the architecture contract:*

$$state(ps(\#ps - 1)) \implies gr \qquad (6)$$

*the time of the last entry corresponds to the duration of the architecture contract:*

$$time(ps(\#ps - 1)) = d \qquad (7)$$

*and for all entries $0 \leq i < \#ps$, such that $rat(ps(i)) = (tg', gr', d')$:*
1. $ps(i)$ *refers only to triggers of the architecture contract or previous proof steps:*

   $\forall j < \#ref(ps(i))\colon$
   $\bigl(\forall k \in ref(ps(i))(j)\colon k < \#tg\bigr) \land \bigl(\forall (k,n) \in ref(ps(i))(j)\colon k < i\bigr)$

   *Note that this implies that $ref(ps(0))$ contains only references to triggers $tg$ of the architecture contract.*

2. *The time points of the referenced entries respect the time points of the triggers of the rationale. Thus, we first introduce a function* $time$ *(for* $\#ref(ps(i)) > 0$*) to return the* relative *time point of a reference:*

$$time(ref(ps(i))(j)) = \begin{cases} time(tg(k)) & \text{if } k \in ref(ps(i))(j) \\ time(ps(k)) & \text{if } (k,n) \in ref(ps(i))(j) \end{cases}$$

*Now we can use this function to formalize the condition (note that by definition 8, $\#ref(ps(i)) = \#tg'$):*

$$\forall j < \#ref(ps(i)) \colon time(ref(ps(i))(j)) = time(ref(ps(i))(0)) + time(tg'(j))$$

*Note that this condition implies that for all $j < \#ref(ps(i))$, the referenced time points of all entries $e \in ref(ps(i))(j)$ is the same.*

3. *The referenced entries imply the corresponding triggers of the rationale:*

$$\forall j < \#ref(ps(i)) \colon \left( \bigwedge_{k \in ref(ps(i))(j)} state(tg(k)) \right) \wedge$$

$$\left( \bigwedge_{(k,n) \in ref(ps(i))(j)} state(ps(k)) \wedge \bigwedge_{(p_i, p_o) \in n} p_i = p_o \right) \implies state(tg'(j))$$

4. *The time of the current entry respects the duration of the rationale (note that the time of $ref(ps(i))(0)$ corresponds to the time point of the first trigger of the rationale):*

$$time(ps(i)) = time(ref(ps(i))(0)) + d'$$

5. *The guarantee of the rationale implies the current state:*

$$gr' \implies state(ps(i))$$

Similar as for contracts, architecture proofs are best expressed graphically using a notation similar to Message Sequence Charts (see fig. 6 for an example).

### 4.1 Verifying Reliable Adder

Table 1 shows an architecture proof for the contract of our running example depicted in fig. 5. It consists of four steps: 0. First, we apply contract `dispatch` of the dispatcher component to trigger 0 of the architecture contract, to obtain a valuation of the dispatcher's output ports at time point 1 with messages $x$ and $y$, respectively. 1. Then, we use connections $(a1i1, do1)$ and $(a1i2, do2)$ to pass messages $x$ and $y$ to the corresponding input ports of Adder1 and apply contract `add1` to obtain a new state for time point 5, in which the output port of Adder1 contains the sum of $x$ and $y$. 2. Similarly, we can use connections $(a2i1, do3)$ and $(a2i2, do4)$ to apply contract `add2` to the architecture state given by step 0, to obtain a new state for time point 4, in which the output port of Adder2 contains the sum of $x$ and $y$. 3. Finally, we can use connections $(mi1, a1o)$ and $(mi2, a2o)$ to pass the calculated sums to the input of the merger component and apply contract merge3 to forward it on its output port. Note that the proof is only valid, since we chose contract merge3 for the merger component. If we had chosen merge1 or merge2, the proof would have violated condition 2 of definition 9.

Table 1: Architecture proof for reliable adder.

| | tp | $\gamma$ | r | rf |
|---|---|---|---|---|
| 0 | 1 | $do1 = x \land do2 = y$ | dispatch | $\{0\}$ |
| 1 | 5 | $a1o = x + y$ | add1 | $\{(0, \{(a1i1, do1), (a1i2, do2)\})\}$ |
| 2 | 4 | $a2o = x + y$ | add2 | $\{(0, \{(a2i1, do3), (a2i2, do4)\})\}$ |
| 3 | 7 | $mo = x + y$ | merge3 | $\{(1, \{(mi1, a1o)\}), (2, \{(mi2, a2o)\})\}$ |

As mentioned above, architecture proofs can also be expressed graphically using a notation similar to Message Sequence Charts. For example, the proof from table 1, could also be expressed graphically as depicted in fig. 6.

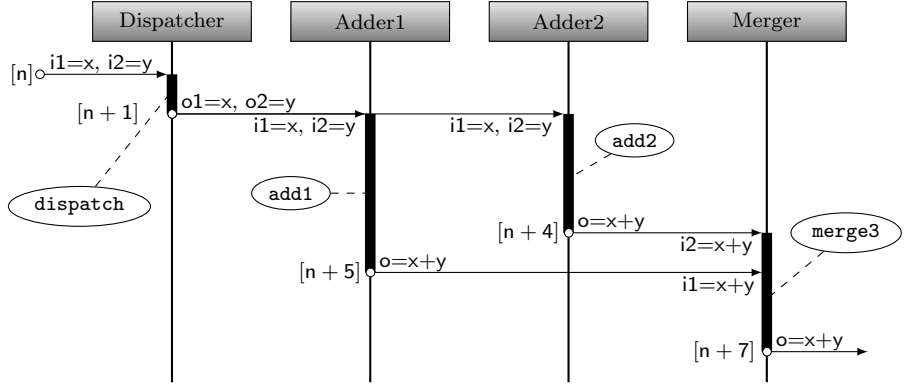

Fig. 6: Architecture proof by means of Message Sequence Chart.

### 4.2 Soundness and Completeness

In the following, we provide two theoretical results for APML. The first one ensures that if we can prove an architecture contract from the contracts of its components using APML, then, an architecture in which the components satisfy the corresponding contracts is indeed guaranteed to satisfy the architecture contract.

**Theorem 1 (Soundness).** *Given an architecture $A = (F, N)$ and corresponding contracts $K_f$ for each interface $f \in F$, such that $\forall f \in F \colon \mathcal{B}_f \models K_f$. If there exists an architecture proof ps for an architecture contract $k = (tg, gr, d)$, we have $_A \otimes \mathcal{B} \models k$.*

*Proof (The full proof is provided in appendix A).* According to definition 7 we have to show that for all $t \in {_A}\otimes\mathcal{B}$ and all $n \in \mathbb{N}$, $\big(\forall j < \#tg \colon t(n + time(tg(j))) \models state(tg(j))\big) \implies t(n+d) \models gr$. Thus, we assume $\forall j < \#tg \colon t(n+time(tg(j))) \models state(tg(j))$ and we show by *complete* induction over the length of the proof sequence that $\forall i < \#ps \colon t(n+time(ps(i))) \models state(ps(i))$. Thus, $t(n+time(ps(\#ps-1))) \models state(ps(\#ps-1))$ and, by eq. (6) and eq. (7), we can conclude $t(n+d) \models gr$. □

The second result guarantees that, whenever the satisfaction of contracts for components of an architecture leads to the satisfaction of a corresponding contract for the architecture, then it is possible to find a corresponding APML proof.

**Theorem 2 (Completeness).** *Given an architecture $A = (F, N)$ and corresponding contracts $K_f$ for each interface $f \in F$. For each architecture contract $k$, such that for all possible behaviors $\mathcal{B}$:*

$$\bigl(\forall f \in F \colon \mathcal{B}_f \models K_f\bigr) \implies {}_A\otimes\mathcal{B} \models k \tag{8}$$

*there exists an architecture proof $ps$ for $k = (tg, gr, d)$.*

*Proof (The full proof is provided in appendix B).* For the proof we construct a "maximal" architecture proof $ps$, according to definition 9, by repeatedly applying all feasible contracts. If we eventually reach an entry such that $state(ps(\#ps - 1)) \implies gr$ and $time(ps(\#ps - 1)) = d$ then we are done. If not, then we build an architecture trace $t \in \overline{(port(F))^\infty}$, such that $\forall j < \#tg \colon t(time(tg(j))) \models state(tg(j))$ and $\forall i < \#ps \colon t(time(ps(i))) \models state(ps(i))$ and $t(d) \not\models gr$ and for all other $n$, we choose $t(n)$, such that the projection to every interface $f \in F$ does not satisfy the assumptions of any contract $K_f$. Now, we can show that $\forall f \in F \colon \mathcal{B}_f \models K_f$ and thus, by eq. (8) we can conclude ${}_A\otimes\mathcal{B} \models k$. Thus, since $\forall j < \#tg \colon t(time(tg(j))) \models state(tg(j))$, we can conclude $t(d) \models gr$ which is a contradiction to $t(d) \not\models gr$. □

---

**Algorithm 1** Mapping an APML proof to a corresponding proof in Isabelle/ISAR

---

**Input:** a proof $ps$ according to definition 9 and a function $toIsabelle$ to convert port predicates
**Output:** a proof in Isabelle/Isar [42]
1: $i = 0$
2: **while** $i < \#ps$ **do**
3:   $(tp, \gamma, r, rf) := ps(i); (tg, gr, d) := r;$
4:   **if** $rf = \langle\rangle$ **then print** "have s" $+ i +$ ": " $+ toIsabelle(\gamma, tp) +$ " by simp" **else**
5:     $i' = 0$
6:     **while** $i' < \#rf$ **do**
7:       **if** $i' > 0$ **then print** "moreover " **end if**
8:       **print** "from "
9:       **for all** $i'' \in rf(i')$ **do print** "a" $+ i'' +$ " " **end for**
10:      **for all** $(i'', n') \in rf(i')$ **do print** "s" $+ i'' +$ " " **end for**
11:      **print** "have " $+ toIsabelle(state(tg(i')), time(tg(i'))) +$ " "
12:      **if** $rf(i') \setminus \mathbb{N} \neq \emptyset$ **then print** "using " **end if**
13:      **for all** $(i'', n') \in rf(i')$ and $(p_i, p_o) \in n'$ **do print** "$p_i\_p_o$ " **end for**
14:      **print** " by simp"
15:      $i'++$
16:     **end while**
17:     **if** $i' = 1$ **then print** "hence " **else**
18:     **if** $i' > 1$ **then print** "ultimately have " **end if**
19:     **print** "s" $+ i +$ ": " $+ toIsabelle(\gamma, tp) +$ " using " $+ r +$ " by blast" **end if**
20:   $i++$
21: **end while**
22: **print** "thus ?thesis by auto"

---

# 5 From APML to Isabelle

To verify soundness of an APML proof, algorithm 1 shows how an APML proof can be mapped to a corresponding Isar proof for the interactive theorem prover Isabelle.

Let us see how the algorithm can be applied to generate an Isar proof for the APML proof of our running example, described in section 4.1. First, we create an Isabelle locale for the architecture as described in fig. 3:

```
locale rsum =
 fixes
  — Dispatcher: di1::nat⇒nat and di2::nat⇒nat
  and do1::nat⇒nat and do2::nat⇒nat and do3::nat⇒nat and do4::nat⇒nat
  — Adder1: and a1i1::nat⇒nat and a1i2::nat⇒nat and a1o::nat⇒nat
  — Adder2: and a2i1::nat⇒nat and a2i2::nat⇒nat and a2o::nat⇒nat
  — Merger: and mi1::nat⇒nat and mi2::nat⇒nat and mo::nat⇒nat
  — Contracts:
 assumes dispatch: ⋀n x y. ⟦di1 n = x; di2 n = y⟧ ⟹
   do1 (n+1) = x ∧ do2 (n+1) = y ∧ do3 (n+1) = x ∧ do4 (n+1) = y
  and add1: ⋀n x y. ⟦a1i1 n = x; a1i2 n = y⟧ ⟹ a1o (n+4) = x + y
  and add2: ⋀n x y. ⟦a2i1 n = x; a2i2 n = y⟧ ⟹ a2o (n+3) = x + y
  and merge1: ⋀n x. ⟦mi1 n = x; mi2 n = x⟧ ⟹ mo (n+2) = x
  and merge2: ⋀n x. ⟦mi1 n = x; mi2 (n+1) = x⟧ ⟹ mo (n+3) = x
  and merge3: ⋀n x. ⟦mi2 n = x; mi1 (n+1) = x⟧ ⟹ mo (n+3) = x
  — Connections
  and do1-a1i1: ⋀n. a1i1 n = do1 n and a1i2-do2: ⋀n. a1i2 n = do2 n
  and do3-a2i1: ⋀n. a2i1 n = do3 n and a2i2-do4: ⋀n. a2i2 n = do4 n
  and a1o-mi1: ⋀n. mi1 n = a1o n and mi2-a2o: ⋀n. mi2 n = a2o n
```

Note that each contract, as presented in fig. 4, results in a corresponding locale assumption. Now, we can create a theorem for the architecture contract described by fig. 5:

```
theorem sum:
 fixes n x y assumes a0: di1 n = x ∧ di2 n = y
 shows mo (n+7) = x + y
```

Finally, we can apply algorithm 1 to create an Isar proof for the theorem from the APML proof described in table 1:

```
proof −
 from a0 have di1 n = x ∧ di2 n = y by auto
 hence s1: do1 (n+1) = x ∧ do2 (n+1) = y ∧ do3 (n+1) = x ∧ do4 (n+1) = y
  using dispatch by blast
 from s1 have a1i1 (n+1) = x ∧ a1i2 (n+1) = y using do1-a1i1 a1i2-do2 by auto
 hence s2: a1o (n+5) = x + y using add1 by blast
 from s1 have a2i1 (n+1) = x ∧ a2i2 (n+1) = y using do3-a2i1 a2i2-do4 by auto
 hence s3: a2o (n+4) = x + y using add2 by blast
 from s2 have mi1 (n+5) = x + y using a1o-mi1 by auto
 moreover from s3 have mi2 (n+4) = x + y using mi2-a2o by auto
 ultimately have mo (n+7) = x+y using merge3 by blast
 thus ?thesis by auto
qed
```

## 6 Modeling Architecture Proofs in FACTUM Studio

To support the development of APML proofs in practice, we implemented the language in FACTUM Studio [31]: an architecture modeling application based on Eclipse/EMF [40]. FACTUM Studio now supports the user in the development of "correct" APML proofs by means of three key features: (i) It analyses the structure of a given APML proof and checks it for syntactical errors. (ii) It uses so-called validators to check for violations of the conditions described in definition 8 and, to a limited extent, also the ones described in definition 9. (iii) The textual development of APML proofs in Xtext [5] is complemented by corresponding graphical notations using Sirius [34]. To support the verification of single proof steps, we implemented algorithm 1 in FACTUM Studio. Thus, after specifying an APML proof, a user can automatically generate a corresponding Isar proof for Isabelle.

Figure 7 depicts the specification of our running example in FACTUM Studio: First, the architecture is specified graphically in terms of interfaces (represented as gray rectangles) and connections between their input (empty circles) and output (filled circles) ports. Then, contracts can be added for each component using a textual notation.

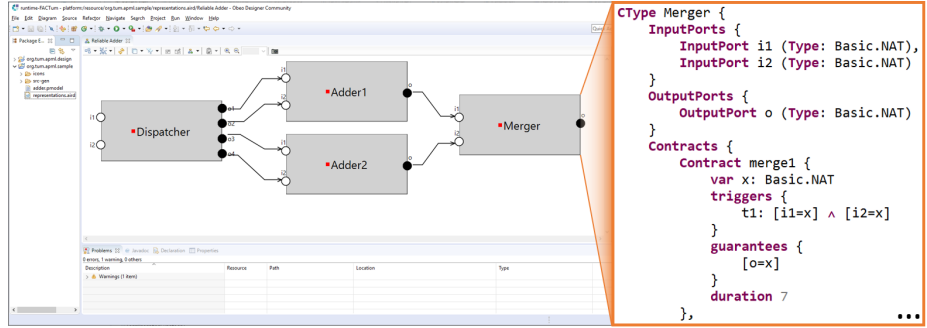

Fig. 7: Specification of reliable adder in FACTUM Studio.

Figure 8 shows how we can verify the adder system using FACTUM Studio's APML implementation: After specifying the contract, we can provide a corresponding proof in terms of a sequence of proof steps as described in definition 8.

As mentioned above, FACTUM Studio performs several checks to ensure consistency of proofs: (i) For each step it checks whether enough rationales are provided (last condition of definition 8). (ii) It verifies that only existing connections are used in the "with" clause (last condition of definition 8). (iii) It ensures that each step only refers to triggers of the contract or previous proof steps (condition 1 of definition 9) (iv) It ensures consistency of the time points of contracts with those of rationales (conditions 2 and 4 of definition 9).

Fig. 8: APML proof in FACTUM Studio.

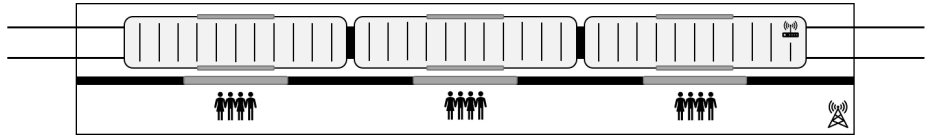
Fig. 9: Automatic Train Control System

## 7 Case Study: Trainguard MT Control System

We evaluated our approach by applying it for the verification of a railway-control system. To investigate whether the approach can indeed be used to apply interactive theorem proving by users not trained in using this technology, verification was performed by a subject with no prior experience with formal methods in general and specifically with the use of interactive theorem provers. The subject holds a Bachelor of Science degree in computer science and has four years of industrial experience as software developer in the domain of business information systems.

*Study Context.* Trainguard MT (TGMT) is an automatic train control system for metros, rapid transit, commuter, and light rail systems currently in development by one of our industrial partners. For the purpose of this case study, we focused on the verification of one key feature of TGMT: controlling of the platform screen doors (PSDs). The situation is depicted by fig. 9: In an effort to make train stations safer for the passengers, modern stations protect their rail zone with a wall (represented by a black line) with several doors (represented by gray lines), the so-called PSDs. When a train arrives at the station, its onboard control unit communicates with the Wayside control unit to control opening/closing of the PSDs.

*Study Setup.* To model the PSD functionality, the company provided four documents as input, which were taken directly from the PSD development: a high-level system requirements specification (59 pages), a more detailed system architecture specification (299 pages), a performance specification (57 pages), and a glossary (42 pages). Based on the documents, we specified a corresponding architecture for the PSD functionality, consisting of 33 components with 36 contracts in total.

*Study Execution.* The subject then verified five architecture contracts: (P1) If the train is moving, the PSDs are closed. (P2) If the train is at standstill and the position of the train doors match the position of the PSDs, the PSDs are opened. (P3) If the train doors open on the right hand side, the platform must be on the right hand side. (P4) If door release is permissive and the train is at standstill, its doors are open. (P5) When the train indicates that the doors are closed, the PSDs are closed.

*Results.* After a study of the architecture and a brief introduction into APML, the subject was able to verify versions of all the properties with no further guidance. Figure 10 depicts the verification effort in terms of APML proof steps required for each property. In total, the subject required roughly 25 working hours to develop the proofs. Sometimes, however, the original contracts needed to be adapted to fit the structure of a contract as defined by definition 6. This was mainly due to two reasons: (i) Some of the contracts required to express state-ments of the form "whenever x happens, after n time points y happens *and in between z holds*". (ii) In addition, some components required contracts combined by disjunction rather than conjunction. Moreover, the proofs for properties P1 and P5 share a common proof sequence of 8 proof steps, which is more than $50\%$ of the steps required to proof the properties. Finally, some of the proof steps could only be discharged by adapting the generated Isabelle script to tune the simplifier and the logical reasoner, respectively.

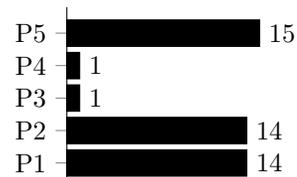

Fig. 10: Number of proof steps for each property.

*Conclusions.* From our results, we derive the following conclusions:

**C1** The approach can be used to support users with no prior experience with formal methods in the development of proofs for interactive theorem provers.
**C2** Some of the properties sometimes required in a practical setting cannot yet be expressed by the notion of contract used in this paper.
**C3** The verification of properties can involve considerable, redundant proof sequences.
**C4** Some initial setup may be required to automatically discharge generated proof steps.

## 8 Related Work

Related work can be found in two different areas: verification of contract-based specifications and interactive verification of component-based systems.

*Verification of contract-based specifications.* Verification of contract-based specifications is not a new topic, at all. First works in this area go back to Pnueli's work about modular model checking [36] and Clarke's work on compositional model checking [14], in which they investigated possibilities to leverage assumptions about a component's environment for the verification of its implementation. Later on, attempts were made to automatically synthesize minimal assumptions from a component's implementation [7,19,20,35]. In all these studies, the focus is mainly on the verification of component implementations under the presence of assumptions about its environment. In contrary, the focus of the work presented in this paper, is on the verification of contract-composition, without even considering component implementations.

Another line of research in this area is work which focuses on the analysis of contract compatibility. In parametric AG reasoning [38], there may even be many assumptions and guarantees for each component. An example of work in this area is Damm et al. [16], in which the authors describe an approach for virtual integration testing based on rich components. Another example is the work of Chilton et al. [11], which provides an approach to AG reasoning for safety properties. Later on, Kugele and Marmsoler [25] provide an approach to check compatibility of AG contracts using SMT solvers [17]. Finally, Broy [8] provides a detailed characterization about compatible assumptions and guarantees for FOCUS [9]. While the work in this line of research also focuses on the verification at the level of contracts, the focus is mainly on the verification of contract compatibility. With our work, however, we focus on a related, though different problem: verification of refinement of contracts.

Work in this area of research which is most closely related to our work is Cimatti's work on OCRA [12,13]. Here, the author proposes an approach based on bounded model checking [6] to verify refinement of LTL-based contracts. With our work, we follow a similar goal by applying an alternative approach based on interactive theorem proving, which is why we actually complement Cimatti's work.

*Interactive Verification Component based Systems.* There exists some work which investigates the application of interactive theorem proving (ITP) for the verification of component-based systems. Fensel and Schnogge [21], for example, apply the KIV interactive theorem prover [37] to verify architectures in the area of knowledge-based systems. More recently, some attempts were made to apply ITP for the verification of architectural connectors: Li and Sun [26], for example, apply the Coq proof assistant [4] to verify connectors specified in Reo [1]. Moreover, there exists work on the formalization of OCL [41] in Isabelle [10]. In addition, Spichkova [39] provides a framework for the verification of component-based systems in Isabelle and Marmsoler [28] extends the work to deal with dynamic reconfiguration. Also the mechanization of UTP [23] in Isabelle, due by Foster et al. [22], belongs to this line of research. However, while most of these works focus on the interactive verification of component based systems, verification is usually done at the level of the prover. With the work presented in this paper, we aim to contribute to this line of research by exploring possibilities and limitations of synthesizing proofs for the provers from architectural descriptions.

*Summary.* Concluding, to the best of our knowledge, this is the first attempt to synthesize proof for an interactive theorem prover for the verification of contract composition from architectural descriptions.

## 9 Conclusion

With this paper, we introduced APML: an architecture proof modeling language. To this end, we first introduced a formal semantics for component contracts as well as architecture contracts. Then, we described APML and showed soundness and completeness of it for the verification of architecture contracts. Moreover, we presented an algorithm to map an APML proof to a corresponding proof in Isabelle/Isar and discussed its implementation in Eclipse/EMF. Finally, we demonstrated the approach by means of a running example and evaluated it by means of a larger case study from the railway domain.

Our theoretical results (Section 4.2) show that APML can indeed be used to specify abstract proofs for the composition of contracts using a notation similar to Message Sequence Charts. Moreover, as indicated by C1 of our case study, it supports users with no prior experience in interactive theorem proving in the development of proofs. Thus, APML indeed contributes to the overall goal of bridging the gap between software architecture and interactive theorem proving. Nevertheless, the case study also revealed some limitations of the approach, which should be addressed by future work: As indicated by C2, future work should investigate possibilities to enhance expressiveness of the contracts. Specifically, the support for until-like contracts and disjunction of contracts should be investigated. Moreover, as indicated by C3, future work should investigate possibilities to support reuse of proof sequences. To this end, a proof step should be allowed to reference to already verified contracts. Finally, as indicated by C4, future work should investigate possibilities to increase automation at the level of interactive theorem proving.

*Acknowledgments.* We would like to thank Simon Foster and Mario Gleirscher for inspiring discussions about APML. Parts of the work on which we report in this paper was funded by the German Federal Ministry of Education and Research (BMBF) under grant no. 01Is16043A.

## A  Proof for Thm. 1

The following diagram depicts an overview of the main concepts and its relationships and marks the proof obligation with a question mark:

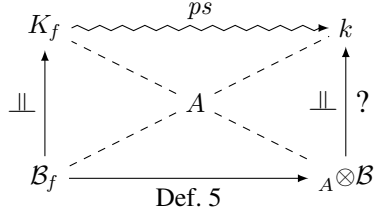

According to Def. 7 we have to show that for all $t \in {}_A \otimes \mathcal{B}$ and all $n \in \mathbb{N}$, $\big(\forall i < \#tg\colon t(n + time(tg(i))) \models state(tg(i))\big) \implies t(n + d) \models gr$.

– Thus, we assume

$$\forall i < \#tg\colon t(n + time(tg(i))) \models state(tg(i)) \qquad (9)$$

and we show by induction that $\forall i < \#ps\colon t(n + time(ps(i))) \models state(ps(i))$.

- Thus, fix $i < \#ps$ and assume $\forall i' < i\colon t(n + time(ps(i'))) \models state(ps(i'))$.
- Let $(tp, \gamma, r, rf) = ps(i)$.
- Since, by assumption, we have $\forall f \in F\colon \mathcal{B}_f \models K_f$, conclude $\mathcal{B}_f \models r$ for some $f \in F$.
- Thus, let $r = (tg', gr', d')$ and by Def. 7 have $\forall t' \in \mathcal{B}_f$, $n' \in \mathbb{N}\colon \big(\forall i'' < \#tg'\colon t'(n' + time(tg'(i''))) \models state(tg'(i''))\big) \implies t'(n' + d') \models gr'$.
- Thus, by Eq. (4) of Def. 5 have $\forall n' \in \mathbb{N}\colon \big(\forall i'' < \#tg'\colon t(n' + time(tg'(i''))) \models state(tg'(i''))\big) \implies t(n' + d') \models gr'$.
- Thus, we show $\forall i'' < \#tg'\colon t(n + time(rf(0)) + time(tg'(i''))) \models state(tg'(i''))$ to conclude $t(n + time(rf(0)) + d') \models gr'$.
  * Thus, let $i'' < \#tg'$ and since by Def. 8 $\#rf = \#tg'$, conclude $i'' < \#rf$.
  * Thus, by condition 3 of Def. 9 have

$$\left( \bigwedge_{i''' \in rf(i'')} state(tg(i''')) \right) \wedge$$

$$\left( \bigwedge_{(i''', n'') \in rf(i'')} state(ps(i''')) \wedge \bigwedge_{(p_i, p_o) \in n''} p_i = p_o \right) \implies state(tg'(i''))$$

  * Thus, we show

$$\forall i''' \in rf(i'')\colon t(n + time(rf(0)) + time(tg'(i''))) \models state(tg(i''')) \qquad (10)$$

and

$$\forall (i''', n'') \in rf(i'')\colon t(n + time(rf(0)) + time(tg'(i''))) \models state(ps(i''')) \\ \wedge \forall (p_i, p_o) \in n''\colon t(n + time(rf(0)) + time(tg'(i''))) \models p_i = p_o \qquad (11)$$

i

to conclude $t(n + time(rf(0)) + time(tg'(i''))) \models state(tg'(i''))$:
- Eq. (10): Let $i''' \in rf(i'')$ and have $i''' < \#tg$ by condition 1 of Def. 9. Thus, have $t(n + time(tg(i'''))) \models state(tg(i'''))$ by Eq. (9). Moreover, since $i''' \in rf(i'')$ have $time(rf(i'')) = time(tg(i'''))$ from the definition of $time$ in condition 2 of Def. 9. Thus, since $rf = ref(ps(i))$ and $i'' < \#ref(ps(i))$ (since $i'' < \#tg'$ and by Def. 8 $\#tg' = \#ref(ps(i))$) have $time(tg(i''')) = time(rf(0)) + time(tg'(i''))$ by condition 2 of Def. 9. Thus, since $t(n + time(tg(i'''))) \models state(tg(i'''))$ conclude $t(n + time(rf(0)) + time(tg'(i''))) \models state(tg(i'''))$.
- Eq. (11): We show that for all $(i''', n'') \in rf(i'')$,

$$t(n + time(rf(0)) + time(tg'(i''))) \models state(ps(i''')) \quad (12)$$

and

$$\forall (p_i, p_o) \in n'' : t(n+time(rf(0))+time(tg'(i''))) \models p_i = p_o \quad (13)$$

Thus, let $(i''', n'') \in rf(i'')$.
- Eq. (12): By condition 1 of Def. 9, $i''' < i$ and thus we can apply the induction hypothesis to have $t(n + time(ps(i'''))) \models state(ps(i'''))$. Moreover, since $(i''', n'') \in rf(i'')$ have $time(ps(i''')) = time(rf(i''))$ from the definition of $time$ in condition 2 of Def. 9. Thus, since $rf = ref(ps(i))$ and $i'' < \#ref(ps(i))$ (since $i'' < \#tg'$ and by Def. 8 $\#tg' = \#ref(ps(i))$) have $time(ps(i''')) = time(rf(0)) + time(tg'(i''))$ by condition 2 of Def. 9. Thus, since $t(n + time(ps(i'''))) \models state(ps(i'''))$ conclude $t(n + time(rf(0)) + time(tg'(i''))) \models state(ps(i'''))$.
- Eq. (13): Let $(p_i, p_o) \in n''$ and conclude $(p_i, p_o) \in N$ from Def. 8. Thus, by Eq. (5) of Def. 5 have $t(n + time(rf(0)) + time(tg'(i'')))(i) = t(n+time(rf(0))+time(tg'(i'')))(o)$. Finally conclude $t(n + time(rf(0)) + time(tg'(i''))) \models p_i = p_o$.
- From $t(n+time(rf(0))+d') \models gr'$ we can conclude $t(n+time(ps(i))) \models gr'$ by condition 4 of Def. 9 (since $rf = ref(ps(i))$).
- Thus, we conclude $t(n + time(ps(i))) \models state(ps(i))$ by condition 5 of Def. 9.
- From $\forall i < \#ps : t(n + time(ps(i))) \models state(ps(i))$ have $t(n + time(ps(\#ps - 1))) \models state(ps(\#ps - 1))$ (since by Def. 9 $\#ps \geq 1$).
- Thus, from Eq. (6) of Def. 9 have $t(n + time(ps(\#ps - 1))) \models gr$.
- Thus, from Eq. (7) of Def. 9 have $t(n + d) \models gr$.

□

# B  Proof for Thm. 2

The proof proceeds in four steps: First, we construct a "maximal" architecture proof. If it is a valid proof for the architecture contract, we are done, otherwise, we derive a contradiction: To this end, we construct an architecture trace which satisfies all contracts but violates the guarantee of the architecture contract.

The following figure depicts the relationship of an architecture proof (consisting of a sequence of proof steps) and an architecture trace (consisting of a sequence of architecture states).

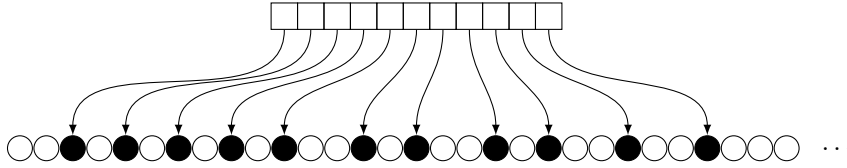

**Constructing the Architecture Proof**

For the proof we construct a "maximal" architecture proof $ps$, according to Def. 9, by repeatedly applying all feasible contracts:
  – closure: for all contracts $(tg', gr', d')$ we have:

$$\forall i < \#tg' \; \exists n: \bigwedge_{\{x \in tg: \; time(x)=n+time(tg'(i))\}} state(x) \wedge \bigwedge_{\{s \in ps: \; time(s)=n+time(tg'(i))\}} state(s) \implies state(tg'(i)) \quad (14)$$

implies
$$\exists i' < \#ps: rat(ps(i')) = c \wedge time(ps(i')) = n + d' \quad (15)$$

  – strongest: for all $i < \#ps$, such that $rat(ps(i)) = (tg', gr', d')$, we have

$$state(ps(i)) \implies gr' \quad (16)$$

If we eventually reach an entry such that $state(ps(\#ps - 1)) \implies gr$ and $time(ps(\#ps - 1)) = d$ then we are done.

**Constructing the Architecture Trace**

If not, then we build an architecture trace $t \in (\overline{port(F)})^\infty$, such that
  – $t$ is the weakest architecture satisfying the architectural assumptions, i.e., for all $n \in \mathbb{N}$, such that $\exists i < \#tg: time(tg(i)) = n$ or $\exists i < \#ps: time(ps(i)) = n$, we have

$$t(n) \models \bigwedge_{\{x \in tg: \; time(x)=n\}} state(x) \wedge \bigwedge_{\{s \in ps: \; time(s)=n\}} state(s) \quad (17)$$

and

$$\forall P \in \Gamma(out(F)) \colon (t(n) \models P) \implies$$

$$\left( \left( \bigwedge_{\{x \in tg \colon time(x)=n\}} state(x) \land \bigwedge_{\{s \in ps \colon time(s)=n\}} state(s) \right) \implies P \right) \quad (18)$$

- for all other $n \in \mathbb{N}$ (where $\neg \exists i < \#tg \colon time(tg(i)) = n$ and $\neg \exists i < \#ps \colon time(ps(i)) = n$) it does not satisfy any assumption:

$$\forall n' \in \mathbb{N},\ (tg', gr', d') \in K_f,\ i < \#tg' \colon \Big(n' + time(tg'(i)) = n\ \land$$

$$\big(\forall i' < \#tg' \colon n' + time(tg'(i')) \neq n \implies t(n' + time(tg'(i'))) \models state(tg'(i')) \big)\Big)$$

$$\implies t(n' + time(tg'(i))) \not\models state(tg'(i))$$

(19)

- $t$ does not satisfy the guarantee at the time point given by the duration:

$$t(d) \not\models gr \quad (20)$$

**Lemma: Application of Contracts in Proofs**

If a contract can be applied to $t$, then it is applied in the proof:

$$\forall n \in \mathbb{N},\ c = (tg', gr', d') \in K_f \colon$$

$$\Big( (\forall i < \#tg' \colon (t(n + time(tg'(i))) \models state(tg'(i))))$$

$$\implies (\exists i' < \#ps \colon rat(ps(i')) = c \land time(ps(i')) = n + d') \Big) \quad (21)$$

- Assume there exists an $n$ and $c = (tg', gr', d') \in K_f$, such that

$$\forall i < \#tg' \colon (t(n + time(tg'(i))) \models state(tg'(i))) \quad (22)$$

- According to Eq. (19), (17), (18), for all $i < \#tg'$

$$t(n + time(tg'(i))) \models$$

$$\bigwedge_{\{x \in tg \colon time(x)=n+time(tg'(i))\}} state(x) \land \bigwedge_{\{s \in ps \colon time(s)=n+time(tg'(i))\}} state(s)$$

(23)

and

$$\forall P \in \Gamma(out(F)) \colon (t(n + time(tg'(i))) \models P) \implies$$

$$\bigwedge_{\{x \in tg \colon time(x)=n+time(tg'(i))\}} state(x) \land \bigwedge_{\{s \in ps \colon time(s)=n+time(tg'(i))\}} state(s)$$

$$\implies P \quad (24)$$

- Thus, since $\forall i < \#tg'\colon t(n + time(tg'(i))) \models state(tg'(i))$ we have for all $i < \#tg'$:

$$\bigwedge_{\{x \in tg\,\colon\, time(x)=n+time(tg'(i))\}} state(x) \wedge \bigwedge_{\{s \in ps\,\colon\, time(s)=n+time(tg'(i))\}} state(s)$$
$$\implies state(tg'(i)) \quad (25)$$

- Thus, since $ps$ is maximal (Eq. (15)), there must exist an entry $i' < \#ps$, such that $rat(ps(i')) = c \wedge time(ps(i')) = n + d'$.

**Deriving the Contradiction**

Now, we can show that $\forall f \in F\colon \mathcal{B}_f \models K_f$.
- Assume $\neg \forall f \in F\colon \mathcal{B}_f \models K_f$.
- Thus, $\exists f \in F\colon \mathcal{B}_f \not\models K_f$ and hence $\exists K = (tg', gr', d') \in K_f\colon t \not\models K$.
- Thus, according to Def. 7 we have $\exists n \in \mathbb{N}\colon \Big(\big(\forall i < \#tg'\colon t(n + time(tg'(i))) \models state(tg'(i))\big) \wedge t(n + d') \not\models gr'\Big)$.
- Thus, according to Eq. (21), $\exists i' < \#ps\colon rat(ps(i')) = K \wedge time(ps(i')) = n+d'$.
- Thus, according to Eq. (17), $t(n+d') \models state(ps(i'))$.
- Moreover, since $rat(ps(i')) = K$ have $state(ps(i')) \implies gr'$ according to Eq. (16).
- Thus, since $t(n+d') \models state(ps(i'))$ conclude $t(n+d') \models gr'$ which is a contradiction to $t(n+d') \not\models gr'$.

Thus, by Eq. (8) we can conclude $_A \otimes \mathcal{B} \models k$. Moreover, by Eq. (17) we can conclude $\forall i < \#tg\colon t(time(tg(i))) \models state(tg(i))$. Thus, since $_A \otimes \mathcal{B} \models k$, according to Def. 7, we can conclude $t(d) \models gr$ which is a contradiction to Eq. (20). □

# C  RAdder Model

```
Pattern TrainguardMT ShortName tgmt {
 DTSpec {
  DT Basic (
   Sort NAT
   Operation add: NAT, NAT => NAT
  )
 }
 CTypes {
  CType Merger {
   InputPorts {
    InputPort i1 (Type: Basic.NAT),
    InputPort i2 (Type: Basic.NAT)
   }
   OutputPorts {
    OutputPort o (Type: Basic.NAT)
   }
   Contracts {
    Contract merge1 {
     var x: Basic.NAT
     triggers {
      t1: [i1=x] /\ [i2=x]
     }
     guarantees {
      [o=x]
     }
     duration 7
    },
    Contract merge2 {
     var x: Basic.NAT
     triggers {
      t1: [i1=x],
      t2: [i2=x] at 1
     }
     guarantees {
      [o=x]
     }
     duration 3
    },
    Contract merge3 {
     var x: Basic.NAT
     triggers {
      t1: [i2=x],
      t2: [i1=x] at 1
```

```
      }
      guarantees {
       [o=x]
      }
      duration 3
     }
    }
  },
  CType Dispatcher {
   InputPorts {
    InputPort i1 (Type: Basic.NAT),
    InputPort i2 (Type: Basic.NAT)
   }
   OutputPorts {
    OutputPort o1 (Type: Basic.NAT),
    OutputPort o2 (Type: Basic.NAT),
    OutputPort o3 (Type: Basic.NAT),
    OutputPort o4 (Type: Basic.NAT)
   }
   Contracts {
    Contract dispatch {
     var x: Basic.NAT,
     var y: Basic.NAT
     triggers {
      t1: [i1=x] /\ [i2=y]
     }
     guarantees {
      [o1=x] /\ [o2=y] /\ [o3=x] /\ [o4=y]
     }
     duration 1
    }
   }
  },
  CType Adder1 {
   InputPorts {
    InputPort i1 (Type: Basic.NAT),
    InputPort i2 (Type: Basic.NAT)
   }
   OutputPorts {
    OutputPort o (Type: Basic.NAT)
   }
   Contracts {
    Contract add1 {
     var x: Basic.NAT,
     var y: Basic.NAT
```

```
      triggers {
       t1: [i1=x] /\ [i2=y]
      }
      guarantees {
       [o=Basic.add[x,y]]
      }
      duration 4
     }
    }
   },
   CType Adder2 {
    InputPorts {
     InputPort i1 (Type: Basic.NAT),
     InputPort i2 (Type: Basic.NAT)
    }
    OutputPorts {
     OutputPort o (Type: Basic.NAT)
    }
    Contracts {
     Contract add2 {
      var x: Basic.NAT,
      var y: Basic.NAT
      triggers {
       t1: [i1=x] /\ [i2=y]
      }
      guarantees {
       [o=Basic.add[x,y]]
      }
      duration 3
     }
    }
   }
  }
  Contracts {
   Contract sum {
    var x: Basic.NAT,
    var y: Basic.NAT
    triggers {
     t1: [Dispatcher.i1=x] /\ [Dispatcher.i2=y]
    }
    guarantees {
     [Merger.o=Basic.add[x,y]]
    }
    duration 7
    proof {
```

```
      s0: at 1
       have [Dispatcher.o1=x] /\ [Dispatcher.o2=y]
       from [
        t1
       ]
       using Dispatcher.dispatch,
      s1: at 5
       have
        [Adder1.o=Basic.add[x,y]]
       from [
        s0 with [
          (Adder1.i1, Dispatcher.o1),
          (Adder1.i2, Dispatcher.o2)
        ]
       ]
       using Adder1.add1,
      s2: at 4
       have
        [Adder2.o=Basic.add[x,y]]
       from [
        s0 with [
          (Adder2.i1, Dispatcher.o3),
          (Adder2.i2, Dispatcher.o4)
        ]
       ]
       using Adder2.add2,
      s3: at 7
       have
        [Merger.o=Basic.add[x,y]]
       from [
        s1 with [
          (Merger.i1, Adder1.o)
        ],
        s2 with [
          (Merger.i2, Adder2.o)
        ]
       ]
       using Merger.merge3
     }
    }
   }
 }
```

# D  TGMT Model

```
Pattern TrainguardMT ShortName tgmt {

 DTSpec {

  DT Status (

   Sort INT

  ),

  DT Int (

   Sort INT

   Operation add: INT, INT => INT,

   sub: INT, INT => INT

  ),

  DT DoorStatus (

   Sort DoorStatus

   Predicate DT_DoorStatus_DoorsOpen: DoorStatus,

   DT_DoorStatus_DoorsVitalClosed: DoorStatus

  ),

  DT Bool (

   Sort BOOLEAN

   Predicate true: BOOLEAN,

   false: BOOLEAN

  ),

  DT Comparison (

   Operation ls: Int.INT, Int.INT => Bool.BOOLEAN,
```

```
  gt: Int.INT, Int.INT => Bool.BOOLEAN
),

DT Command (

 Sort STRING

),

DT DoorReleaseCommand (

 Sort DoorReleaseCommand

 Predicate DT_DoorReleaseCommand_Right:
    DoorReleaseCommand,

 DT_DoorReleaseCommand_Left: DoorReleaseCommand,

 DT_DoorReleaseCommand_Both: DoorReleaseCommand,

 DT_DoorReleaseCommand_Deactivate: DoorReleaseCommand

),

DT DoorOpenStrategy (

 Sort DoorOpenStrategy

 Predicate DT_DoorOpenStrategy_Right: DoorOpenStrategy,

 DT_DoorOpenStrategy_Left: DoorOpenStrategy

),

DT DoorReleaseStatus (

 Sort DoorReleaseStatus

 Predicate DT_DoorReleaseStatus_Released:
    DoorReleaseStatus,

 DT_DoorReleaseStatus_Locked: DoorReleaseStatus
```

),

**DT** DoorStatus (

 **Sort** DoorStatus

 **Predicate** DT_DoorStatus_Opened: DoorStatus,

 DT_DoorStatus_Closed: DoorStatus

),

**DT** MovingStatus (

 **Sort** MovingStatus

 **Predicate** DT_TrainMovingStatus_Moving: MovingStatus,

 DT_TrainMovingStatus_StandStill: MovingStatus

),

**DT** DoorCommand (

 **Sort** DoorCommand

 **Predicate** DT_DoorCommand_Open: DoorCommand,

 DT_DoorCommand_Close: DoorCommand

),

**DT** TrainControlLevel (

 **Sort** TrainControlLevel

 **Predicate** DT_TrainControlLevel_CTC: TrainControlLevel,

 DT_TrainControlLevel_ITC: TrainControlLevel

),

**DT** PSD_Authorization (

 **Sort** PSD_Authorization

```
    Predicate DT_PSD_Authorization_Authorize:
        PSD_Authorization,

    DT_PSD_Authorization_Not_Authorize: PSD_Authorization

  ),

  DT PlatformPSDStatus (

    Sort PlatformPSDStatus

    Predicate DT_PlatformPSDStatus_Has_PSD:
        PlatformPSDStatus,

    DT_PlatformPSDStatus_Has_Not_PSD: PlatformPSDStatus

  ),

  DT StoppingWindow (

    Sort StoppingWindow

    Predicate DT_StoppingWindow_WithinStoppingWindow:
        StoppingWindow,

    DT_StoppingWindow_OutsideStoppingWindow:
        StoppingWindow

  ),

  DT Time (

    Sort INT

  ),

  DT Authority (

    Sort STRING

  )
}
```

```
CTypes {

 /*

 * OBCU ATO Components

 */

 CType OBCU_ATO_In_Converter {

  InputPorts {

   InputPort AB_I_Door_Operating_Mode (Type: DoorStatus.
      DoorStatus),

   InputPort AB_I_Premissive_Door_Button (Type: Bool.
      BOOLEAN),

   InputPort CAB_I_Door_Command (Type: DoorCommand.
      DoorCommand),

   InputPort CL_I_Door_Closed_Indication (Type:
      DoorStatus.DoorStatus)

  }

  OutputPorts {

   OutputPort doorStatus (Type: Status.INT),

   OutputPort cabDoorCommand (Type: DoorCommand.
      DoorCommand),

   OutputPort cabPremissiveDoorButton (Type: Bool.
      BOOLEAN),

   OutputPort cabPremissiveDoorMode (Type: Status.INT)

  }

 },

 CType OBCU_ATO_Door_Mode_Controller {

  InputPorts {
```

```
    InputPort trainOperationMode (Type: Status.INT),

    InputPort platformPSDStatus (Type: Status.INT),

    InputPort trainControlLevel(Type: Status.INT),

    InputPort cabPremissiveDoorButton (Type: Bool.BOOLEAN
       ),

    InputPort cabPremissiveDoorMode (Type: Status.INT)

  }

  OutputPorts {

    OutputPort doorMode (Type: Status.INT)

  }

},

CType OBCU_ATO_Door_Open_Controller {

  InputPorts {

    InputPort doorMode (Type: Status.INT),

    InputPort dwellTimeElapsed (Type: Bool.BOOLEAN),

    InputPort releaseForcedByPermissiveDoorButton (Type:
       Bool.BOOLEAN),

    InputPort doorReleaseStatus (Type: DoorReleaseStatus.
       DoorReleaseStatus),

    InputPort trainControlLevel (Type: Status.INT)

  }

  OutputPorts {

    OutputPort doorOpenCommand (Type: DoorCommand.
       DoorCommand)
```

```
    }

    Contracts {

     Contract doorsOpened {

      triggers {

        doorReleased: DoorReleaseStatus.
            DT_DoorReleaseStatus_Released[doorReleaseStatus
            ],

        timeNotElapsed: Bool.false[dwellTimeElapsed] at 1

      }

      guarantees {

        DoorCommand.DT_DoorCommand_Open[doorOpenCommand]

      }

      duration 2

     },

     Contract doorsClosed {

      triggers {

        doorReleased: DoorReleaseStatus.
            DT_DoorReleaseStatus_Locked[doorReleaseStatus]

      }

      guarantees {

        DoorCommand.DT_DoorCommand_Close[doorOpenCommand]

      }

      duration 1

     }
```

```
   }
  },

  CType OBCU_ATO_Train_Door_Supervision_Mode_Controller {

   InputPorts {

    InputPort movingStatus (Type: MovingStatus.
       MovingStatus),

    InputPort doorReleaseStatus (Type: Time.INT),

    InputPort cabPremissiveDoorButton (Type: Bool.BOOLEAN
       ),

    InputPort trainControlLevel (Type: Status.INT),

    InputPort doorStatus (Type: Status.INT)

   }

   OutputPorts {

    OutputPort trainSupervision (Type: Bool.BOOLEAN)

   }

  },

  CType OBCU_ATO_Train_Door_Supervision_Controller {

   InputPorts {

    InputPort trainSupervision (Type: Bool.BOOLEAN),

    InputPort trainControlLevel (Type: Status.INT),

    InputPort doorStatus (Type: Status.INT)

   }

   OutputPorts {

    OutputPort trainMovement (Type: Bool.BOOLEAN)
```

```
   }

  },

  CType OBCU_ATO_PSD_Door_Controller {

   InputPorts {

    InputPort openDoorCommand (Type: DoorCommand.
       DoorCommand),

    InputPort cabDoorCommand (Type: DoorCommand.
       DoorCommand),

    InputPort trainControlLevel (Type: TrainControlLevel.
       TrainControlLevel)

   }

   OutputPorts {

    OutputPort psdDoorCommand (Type: DoorCommand.
       DoorCommand)

   }

   Contracts {

    Contract forwardDoorCommand {

     triggers {

       controlLevelCTC: TrainControlLevel.
          DT_TrainControlLevel_CTC[trainControlLevel]

     }

     guarantees {

      [psdDoorCommand = openDoorCommand]

     }

     duration 1
```

```
      }
    }
  },
  CType OBCU_ATO_PSD_Door_Command_State {
    InputPorts {
      InputPort openDoorCommand (Type: DoorCommand.
          DoorCommand)
    }
    OutputPorts {
      OutputPort psdDoorCommandState (Type: DoorCommand.
          DoorCommand)
    }
    Contracts {
      Contract forwardingCommand {
        guarantees {
          [psdDoorCommandState = openDoorCommand]
        }
        duration 1
      }
    }
  },
  CType OBCU_ATO_Telegram_Builder {
    InputPorts {
```

```
   InputPort psdDoorCommandState (Type: DoorCommand.
      DoorCommand)

  }

  OutputPorts {

   OutputPort OBCU_WCU_PlatformDoor (Type: DoorCommand.
      DoorCommand)

  }

  Contracts {

   Contract forwardingCommand {

    guarantees {

     [OBCU_WCU_PlatformDoor = psdDoorCommandState]

    }

    duration 1

   }

  }

 },

 CType OBCU_ATO_Out_Converter {

  InputPorts {

   InputPort openDoorCommand (Type: DoorCommand.
      DoorCommand)

  }

  OutputPorts {

   OutputPort TCL_O_Door_Opening_Closing (Type: Status.
      INT)

  }
```

```
        },

/*

* OBCU ITF Components

*/

CType OBCU_ITF_WWA_Telegram_Extractor {

 InputPorts {

  InputPort WCU_OBCU_MovementAuthority (Type: Authority
      .STRING)

 }

 OutputPorts {

  OutputPort rauz (Type: Status.INT)

 }

},

CType OBCU_ITF_Out_Convertor {

 InputPorts {

  InputPort doorStatusHMI (Type: Status.INT),

  InputPort psdStatusHMI (Type: Status.INT),

  InputPort currentDoorModeHMI (Type: Status.INT),

  InputPort doorReleaseStatusHMI (Type: Status.INT)

 }
```

```
    OutputPorts {

     OutputPort HMI_O_Train_Door_Release (Type: Status.INT
        ),

     OutputPort HMI_O_Train_Door_Mode (Type: Status.INT),

     OutputPort HMI_O_PSD (Type: Status.INT),

     OutputPort HMI_O_Train_Door_Status (Type: Status.INT)

    }

  },

  CType OBCU_ITF_HMI_Status_Function {

    InputPorts {

     InputPort rauz (Type: Status.INT),

     InputPort doorReleaseStatus (Type: Status.INT),

     InputPort trainControlLevel (Type: Status.INT),

     InputPort doorMode (Type: Status.INT),

     InputPort doorStatus (Type: Status.INT)

    }

    OutputPorts {

     OutputPort doorStatusHMI (Type: Status.INT),

     OutputPort psdStatusHMI (Type: Status.INT),

     OutputPort currentDoorModeHMI (Type: Status.INT),

     OutputPort doorReleaseStatusHMI (Type: Status.INT)

    }

  },
```

```
CType OBCU_ITF_FWD_Movement_Authority {

 InputPorts {

  InputPort WCU_OBCU_MovementAuthority (Type: Authority
     .STRING)

 }

 OutputPorts {

  OutputPort WCU_OBCU_MovementAuthority (Type:
     Authority.STRING)

 }

},

CType OBCU_ITF_FWD_Position_Report {

 InputPorts {

  InputPort WCU_OBCU_PositionReport (Type:
     PSD_Authorization.PSD_Authorization)

 }

 OutputPorts {

  OutputPort WCU_OBCU_PositionReport_Out (Type:
     PSD_Authorization.PSD_Authorization)

 }

 Contracts {

  Contract forwardAuthorization {

   guarantees {

    [WCU_OBCU_PositionReport_Out =
       WCU_OBCU_PositionReport]

   }
```

```
      duration 1
    }
  }
},
CType OBCU_ITF_FWD_Platform_Door {
  InputPorts {
    InputPort WCU_OBCU_PlatformDoor (Type: DoorCommand.
       DoorCommand)
  }
  OutputPorts {
    OutputPort WCU_OBCU_PlatformDoor_Out (Type:
       DoorCommand.DoorCommand)
  }
  Contracts {
    Contract forwardingCommand {
      guarantees {
        [WCU_OBCU_PlatformDoor_Out = WCU_OBCU_PlatformDoor]
      }
      duration 1
    }
  }
},
```

```
/*
 * OBCU ATP Components
 */

CType OBCU_ATP_In_Converter {

 InputPorts {

  InputPort TCL_I_Door_Closed_Indication (Type:
     DoorStatus.DoorStatus),

  InputPort TCL_I_Door_Closed_Button (Type: Bool.
     BOOLEAN)

 }

 OutputPorts {

  OutputPort trainControlLevel (Type: Status.INT),

  OutputPort cabPremissiveDoorButton (Type: Bool.
     BOOLEAN),

  OutputPort doorStatus (Type: DoorStatus.DoorStatus)

 }

 Contracts {

  Contract forwardDoorStatus {

   guarantees {

    [doorStatus = TCL_I_Door_Closed_Indication]

   }

   duration 1

  }
```

```
   }
  },
  CType OBCU_ATP_Out_Converter {
    InputPorts {
      InputPort propulsion (Type: Bool.BOOLEAN),
      InputPort doorReleaseCommand (Type:
          DoorReleaseCommand.DoorReleaseCommand)
    }
    OutputPorts {
      OutputPort TCL_O_Propulsion_Release (Type: Bool.
          BOOLEAN),
      OutputPort TCL_O_Door_Release (Type: Bool.BOOLEAN)
    }
    Contracts {
      Contract doorsReleased {
        triggers {
          doorsReleased: DoorReleaseCommand.
              DT_DoorReleaseCommand_Left[doorReleaseCommand]
            \/ DoorReleaseCommand.DT_DoorReleaseCommand_Right[
              doorReleaseCommand]
            \/ DoorReleaseCommand.DT_DoorReleaseCommand_Both[
              doorReleaseCommand]
        }
        guarantees {
          Bool.true[TCL_O_Door_Release]
```

```
    }

   duration 1

  },

  Contract doorsNotReleased {

   triggers {

    doorsReleased: DoorReleaseCommand.
        DT_DoorReleaseCommand_Deactivate[
        doorReleaseCommand]

   }

   guarantees {

    Bool.true[TCL_O_Door_Release]

   }

   duration 1

  }

 }

},

CType OBCU_ATP_WMA_Telegram_Extractor{

 InputPorts {

  InputPort WCU_OBCU_MovementAuthority_out (Type:
      Authority.STRING)

 }

 OutputPorts {

  OutputPort rauz (Type: Status.INT)

 }
```

```
            },
            CType OBCU_ATP_PSD_Authorization_State_Tracker {
                InputPorts {
                    InputPort authorizePSDOpening (Type:
                        PSD_Authorization.PSD_Authorization)
                }
                OutputPorts {
                    OutputPort psdAuthorizationState (Type:
                        PSD_Authorization.PSD_Authorization)
                }
                Contracts {
                    Contract forwardAuthorization {
                        guarantees {
                            [psdAuthorizationState = authorizePSDOpening]
                        }
                        duration 1
                    }
                }
            },
            CType OBCU_ATP_PSD_Authorization_Controller {
                InputPorts {
                    InputPort trainControlLevel (Type: TrainControlLevel.
                        TrainControlLevel),
                    InputPort doorReleaseStatus (Type: DoorReleaseStatus.
                        DoorReleaseStatus)
```

```
    }

  OutputPorts {

   OutputPort authorizePSDOpening (Type:
       PSD_Authorization.PSD_Authorization)

  }

  Contracts {

   Contract notAuthorized {

    triggers {

      trainControlLevelCTC: TrainControlLevel.
         DT_TrainControlLevel_CTC[trainControlLevel],

      doorReleaseStatusLocked: DoorReleaseStatus.
         DT_DoorReleaseStatus_Locked[doorReleaseStatus]
         at 1

    }

    guarantees {

      PSD_Authorization.
         DT_PSD_Authorization_Not_Authorize[
         authorizePSDOpening]

    }

    duration 2

   },

   Contract authorized {

    triggers {

      trainControlLevelCTC: TrainControlLevel.
         DT_TrainControlLevel_ITC[trainControlLevel],
```

```
        doorReleaseStatusLocked: DoorReleaseStatus.
            DT_DoorReleaseStatus_Released[doorReleaseStatus]
             at 1

      }

      guarantees {

        PSD_Authorization.DT_PSD_Authorization_Authorize[
            authorizePSDOpening]

      }

      duration 2

    }

  }

},

CType OBCU_ATP_Telegram_Builder {

  InputPorts {

    InputPort psdAuthorizationState (Type:
        PSD_Authorization.PSD_Authorization)

  }

  OutputPorts {

    OutputPort OBCU_WCU_PositionReport (Type:
        PSD_Authorization.PSD_Authorization)

  }

  Contracts {

    Contract forwardAuthorization {

      guarantees {

        [OBCU_WCU_PositionReport = psdAuthorizationState]
```

```
      }
     duration 1
    }
   }
  },
  CType OBCU_ATP_Propulson_Function {
   InputPorts {
    InputPort doorStatus (Type: Status.INT),
    InputPort trainControlLevel (Type: Status.INT),
    InputPort rauz (Type: Status.INT),
    InputPort authorizePSDOpening (Type: Bool.BOOLEAN)
   }
   OutputPorts {
    OutputPort propulsion (Type: Bool.BOOLEAN)
   }
  },
  CType OBCU_ATP_Door_Release_Controller {
   InputPorts {
    InputPort doorStatus (Type: DoorStatus.DoorStatus),
    InputPort trainControlLevel (Type: Status.INT),
    InputPort trainOperationMode (Type: Status.INT),
    InputPort platformPSDStatus (Type: Status.INT),
```

```
    InputPort cabPremissiveDoorButton (Type: Bool.BOOLEAN
       ),

    InputPort berthed (Type: Status.INT),

    InputPort stoppingWindowMinus (Type: Bool.BOOLEAN),

    InputPort stoppingPoint (Type: Bool.BOOLEAN),

    InputPort stopingWindowPlus (Type: Bool.BOOLEAN),

    InputPort trainPosition (Type: Int.INT),

    InputPort movingStatus (Type: MovingStatus.
       MovingStatus),

    InputPort platformSide (Type: DoorOpenStrategy.
       DoorOpenStrategy)

  }

  OutputPorts {

    OutputPort doorReleaseCommand (Type:
       DoorReleaseCommand.DoorReleaseCommand),

    OutputPort doorReleaseStatus (Type: DoorReleaseStatus
       .DoorReleaseStatus),

    OutputPort releaseForcedByPermissiveDoorButton (Type:
        Bool.BOOLEAN)

  }

  Contracts {

    Contract openingLeftDoorSide {

      triggers {

        trigger1: DoorOpenStrategy.DT_DoorOpenStrategy_Left
           [platformSide]

      }
```

```
    guarantees {

      DoorReleaseCommand.DT_DoorReleaseCommand_Left[
         doorReleaseCommand]

    }

    duration 1

  },

  Contract openingRightDoorSide {

    triggers {

      trigger1: DoorOpenStrategy.
         DT_DoorOpenStrategy_Right[platformSide]

    }

    guarantees {

      DoorReleaseCommand.DT_DoorReleaseCommand_Right[
         doorReleaseCommand]

    }

    duration 1

  },

  Contract doorsReleased {

    var trainInStoppingWindow: Bool.BOOLEAN,

    var premissiveRelease: Bool.BOOLEAN

    triggers {

      standStill: MovingStatus.
         DT_TrainMovingStatus_StandStill[movingStatus],
```

```
    trainInStoppingPosition: Bool.true[
        trainInStoppingWindow] at 1,

    premissiveRelease: Bool.false[premissiveRelease] at
        2

  }

  guarantees {

    DoorReleaseStatus.DT_DoorReleaseStatus_Released[
        doorReleaseStatus]

  }

  duration 2

},

Contract doorsNotReleasedWhileMoving {

  triggers {

    moving: MovingStatus.DT_TrainMovingStatus_Moving[
        movingStatus]

  }

  guarantees {

    DoorReleaseStatus.DT_DoorReleaseStatus_Locked[
        doorReleaseStatus] /\

    DoorReleaseCommand.DT_DoorReleaseCommand_Deactivate
        [doorReleaseCommand]

  }

  duration 1

},

Contract doorsClosedWhenClosedIndication {

  triggers {
```

```
        closedIndication: DoorStatus.DT_DoorStatus_Closed[
            doorStatus]

      }

      guarantees {

        DoorReleaseStatus.DT_DoorReleaseStatus_Locked[
            doorReleaseStatus] /\

        DoorReleaseCommand.DT_DoorReleaseCommand_Deactivate
            [doorReleaseCommand]

      }

      duration 1

    },

    Contract doorsReleasedIfStandstillAndInStoppingWindow
        {

      triggers {

        standstill: MovingStatus.
            DT_TrainMovingStatus_StandStill[movingStatus],

        hasPSD: PlatformPSDStatus.
            DT_PlatformPSDStatus_Has_PSD[platformPSDStatus]
            at 1,

        inStopingPosition: Bool.true[Comparison.ls[Int.sub[
            stoppingPoint, stoppingWindowMinus],
            trainPosition]] /\

        Bool.true[Comparison.gt[Int.sub[stoppingPoint,
            stopingWindowPlus], trainPosition]] at 2

      }

      guarantees {

        DoorReleaseStatus.DT_DoorReleaseStatus_Released[
            doorReleaseStatus]
```

```
      }
     duration 3
    }
  }
},

/*
 * Wayside Components
 */
CType Wayside_In_Converter {
 InputPorts {
  InputPort PSD_I_PSD_Overwrite (Type: Bool.BOOLEAN),
  InputPort PSD_I_PSD_Closed (Type: Bool.BOOLEAN),
  InputPort ATS_I_Select_Door_Open_Strategy (Type: Status.INT)
 }
 OutputPorts {
  OutputPort selectedDoorStrategy (Type: Status.INT),
  OutputPort psdClosed (Type: Status.INT),
  OutputPort psdOverwrite (Type: Status.INT)
 }
 Contracts {
  Contract selectedDoorStrategy {
```

```
    guarantees {

      [selectedDoorStrategy =
          ATS_I_Select_Door_Open_Strategy]

    }

    duration 1

   },

   Contract psdClosed {

    triggers {

     trigger1: Bool.true[PSD_I_PSD_Closed]

    }

    guarantees {

     DoorStatus.DT_DoorStatus_DoorsVitalClosed[psdClosed
        ]

    }

    duration 1

   }

  }

 },

 CType Wayside_OPD_Telegram_Extractor {

  InputPorts {

   InputPort OBCU_WCU_PlatformDoor (Type: DoorCommand.
      DoorCommand)

  }

  OutputPorts {
```
<hr />

<hr />

<hr />

```
   OutputPort psdOpenCommand (Type: DoorCommand.
       DoorCommand)

  }

  Contracts {

   Contract forwardCommand {

    guarantees {

     [psdOpenCommand = OBCU_WCU_PlatformDoor]

    }

    duration 1

   }

  }

 },

 CType Wayside_OPR_Telegram_Extractor {

  InputPorts {

   InputPort OBCU_WCU_PositionReport (Type:
       PSD_Authorization.PSD_Authorization)

  }

  OutputPorts {

   OutputPort doorOpenAuthorization (Type:
       PSD_Authorization.PSD_Authorization)

  }

  Contracts {

   Contract forwardAuthorization {

    guarantees {
```

```
        [doorOpenAuthorization = OBCU_WCU_PositionReport]
      }
      duration 1
    }
  }
},
CType Wayside_PSD_Control_Function {
  InputPorts {
    InputPort trainControlLevel (Type: TrainControlLevel.
       TrainControlLevel),
    InputPort doorOpenAuthorization (Type:
       PSD_Authorization.PSD_Authorization),
    InputPort psdOpenCommand (Type: DoorCommand.
       DoorCommand)
  }
  OutputPorts {
    OutputPort psdOpen (Type: Bool.BOOLEAN)
  }
  Contracts {
    Contract doorOpen {
      triggers {
        trainControlLevelITC: TrainControlLevel.
           DT_TrainControlLevel_ITC[trainControlLevel],
```

```
    authorized: PSD_Authorization.
        DT_PSD_Authorization_Authorize[
        doorOpenAuthorization] at 1,

    psdOpenCommand: DoorCommand.DT_DoorCommand_Open[
        psdOpenCommand] at 2

 }

 guarantees {

  Bool.true[psdOpen]

 }

 duration 3

},

Contract doorClosedCTC {

 triggers {

   trainControlLevelITC: TrainControlLevel.
       DT_TrainControlLevel_CTC[trainControlLevel]

 }

 guarantees {

   Bool.false[psdOpen]

 }

 duration 1

},

Contract doorClosedNotAuthorized {

 triggers {

   notAuthorized: PSD_Authorization.
       DT_PSD_Authorization_Not_Authorize[
       doorOpenAuthorization]
```

```
  }

 guarantees {

  Bool.false[psdOpen]

 }

 duration 1

},

Contract doorClosed {

 triggers {

  notAuthorized: PSD_Authorization.
     DT_PSD_Authorization_Not_Authorize[
     doorOpenAuthorization],

  psdCloseCommand: DoorCommand.DT_DoorCommand_Close[
     psdOpenCommand] at 1

 }

 guarantees {

  Bool.false[psdOpen]

 }

 duration 2

},

Contract doorClosedCommandClose {

 triggers {

  psdCloseCommand: DoorCommand.DT_DoorCommand_Close[
     psdOpenCommand]

 }
```

```
       guarantees {
        Bool.false[psdOpen]
        }
       duration 1
      }
     }
   },
   CType Wayside_RAUZ_Control_Function {
    InputPorts {
     InputPort psdOverwrite (Type: Status.INT),
     InputPort trainControlLevel (Type: Status.INT),
     InputPort psdClosed (Type: Bool.BOOLEAN)
    }
    OutputPorts {
     OutputPort rauz (Type: Status.INT),
     OutputPort zoneBlockedATS (Type: Bool.BOOLEAN)
    }
   },
   CType Wayside_Door_Strategy_Function {
    InputPorts {
     InputPort selectedDoorStrategy (Type: Status.INT),
     InputPort defaultDoorStrategy (Type: Status.INT),
     InputPort trainControlLevel (Type: Status.INT)
```

```
  }
  OutputPorts {
   OutputPort currentDoorStrategy (Type: Status.INT)
  }
 },
 CType Wayside_ATS_Status_Function{
  InputPorts {
   InputPort currentDoorStrategy (Type: Status.INT),
   InputPort trainControlLevel (Type: Status.INT)
  }
  OutputPorts {
   OutputPort psdStatusATS (Type: Status.INT),
   OutputPort psdAuthorizationATS (Type: Authority.
      STRING),
   OutputPort psdRequestATS (Type: Status.INT),
   OutputPort doorStrategyStatusToATS (Type: Status.INT)
  }
 },
 CType Wayside_PSD_Door_Command_State_Tracker {
  InputPorts {
   InputPort rauz (Type: Status.INT)
  }
  OutputPorts {
```

```
      OutputPort rauzState (Type: Status.INT)
   }
 },
 CType Wayside_TelegramBuilder {
  InputPorts {
   InputPort rausState (Type: Status.INT)
  }
  OutputPorts {
   OutputPort WCU_OBCU_MovementAuthority (Type:
      Authority.STRING)
  }
 },
 CType Wayside_Out_Converter {
  InputPorts {
   InputPort psdOpen (Type: Bool.BOOLEAN),
   InputPort psdStatusATS (Type: Status.INT),
   InputPort psdAuthorizationATS (Type: Authority.STRING
      ),
   InputPort psdRequestATS (Type: Status.INT),
   InputPort doorStrategyStatusToATS (Type: Status.INT)
  }
  OutputPorts {
   OutputPort PSD_O_PSD_Close (Type: Bool.BOOLEAN),
```

```
    OutputPort PSD_O_PSD_Open (Type: Bool.BOOLEAN),

    OutputPort ATS_O_PSD_Authorization (Type: Authority.
       STRING),

    OutputPort ATS_O_PSD_Status (Type: Status.INT),

    OutputPort ATS_O_PSD_Request (Type: Status.INT),

    OutputPort ATS_O_Door_Strategy (Type: Status.INT)
 }

 Contracts {

  Contract psdOpen {

   triggers {

    open: Bool.true[psdOpen]

   }

   guarantees {

    Bool.true[PSD_O_PSD_Open] /\ Bool.false[
       PSD_O_PSD_Close]

   }

   duration 1

  },

  Contract psdClose {

   triggers {

    open: Bool.false[psdOpen]

   }

   guarantees {
```

```
        Bool.false[PSD_O_PSD_Open] /\ Bool.true[
            PSD_O_PSD_Close]

      }

     duration 1

    }

   }

  }

 }

Contracts {

 // 1) If the train is moving, the PSDs are closed.

  Contract PSDAreClosedWhenTrainIsMoving {

   triggers {

    trigger1: MovingStatus.DT_TrainMovingStatus_Moving[
        OBCU_ATP_Door_Release_Controller.movingStatus]

   }

   guarantees {

    Bool.true[Wayside_Out_Converter.PSD_O_PSD_Close] /\
        Bool.false[Wayside_Out_Converter.PSD_O_PSD_Open]

   }

   duration 13

   proof {

    step1:

    at 3
```

**have**

```
DoorReleaseCommand.DT_DoorReleaseCommand_Deactivate[
    OBCU_ATP_Door_Release_Controller.
    doorReleaseCommand] /\

DoorReleaseStatus.DT_DoorReleaseStatus_Locked[
    OBCU_ATP_Door_Release_Controller.doorReleaseStatus
    ]
```

**from** [

```
trigger1
```

]

**using**

```
OBCU_ATP_Door_Release_Controller.
    doorsNotReleasedWhileMoving,

step2:
```

**at** 4

**have**

```
PSD_Authorization.DT_PSD_Authorization_Not_Authorize[
    OBCU_ATP_PSD_Authorization_Controller.
    authorizePSDOpening]
```

**from** [

```
step1 with [

(OBCU_ATP_PSD_Authorization_Controller.
    doorReleaseStatus,
    OBCU_ATP_Door_Release_Controller.doorReleaseStatus
    )

]

]
```

```
          using
          OBCU_ATP_PSD_Authorization_Controller.notAuthorized,
step3:
    at 5
    have
        PSD_Authorization.DT_PSD_Authorization_Not_Authorize[
            OBCU_ATP_PSD_Authorization_State_Tracker.
            psdAuthorizationState]
    from [
        step2 with [
            (OBCU_ATP_PSD_Authorization_State_Tracker.
                authorizePSDOpening,
             OBCU_ATP_PSD_Authorization_Controller.
                authorizePSDOpening)
        ]
    ]
          using
          OBCU_ATP_PSD_Authorization_State_Tracker.
             forwardAuthorization,
step4:
    at 6
    have
        PSD_Authorization.DT_PSD_Authorization_Not_Authorize[
            OBCU_ATP_Telegram_Builder.OBCU_WCU_PositionReport]
    from [
        step3 with [
```

```
            (OBCU_ATP_Telegram_Builder.psdAuthorizationState,
                OBCU_ATP_PSD_Authorization_State_Tracker.
                psdAuthorizationState)
        ]
    ]
    using
    OBCU_ATP_Telegram_Builder.forwardAuthorization,
step5:
    at 4
    have
    DoorCommand.DT_DoorCommand_Close[
        OBCU_ATO_Door_Open_Controller.doorOpenCommand]
    from [
    step1 with [
        (OBCU_ATO_Door_Open_Controller.doorReleaseStatus,
            OBCU_ATP_Door_Release_Controller.doorReleaseStatus
            )
    ]
    ]
    using
    OBCU_ATO_Door_Open_Controller.doorsClosed,
step6:
    at 5
    have
    DoorCommand.DT_DoorCommand_Close[
        OBCU_ATO_PSD_Door_Controller.psdDoorCommand]
```

**from** [

step5 with [

(OBCU_ATO_PSD_Door_Controller.psdDoorCommand,
    OBCU_ATO_Door_Open_Controller.doorReleaseStatus)

]

]

**using**

OBCU_ATO_PSD_Door_Controller.forwardDoorCommand,

step7:

**at** 6

**have**

DoorCommand.DT_DoorCommand_Close[
    OBCU_ATO_PSD_Door_Command_State.
    psdDoorCommandState]

**from** [

step6 with [

(OBCU_ATO_PSD_Door_Command_State.openDoorCommand,
    OBCU_ATO_PSD_Door_Controller.psdDoorCommand)

]

]

**using**

OBCU_ATO_PSD_Door_Command_State.forwardingCommand,

step8:

**at** 7

**have**

```
DoorCommand.DT_DoorCommand_Close[
    OBCU_ATO_Telegram_Builder.OBCU_WCU_PlatformDoor]
```

**from** [

```
step7 with [

(OBCU_ATO_Telegram_Builder.psdDoorCommandState,
    OBCU_ATO_PSD_Door_Command_State.
    psdDoorCommandState)

]

]
```

**using**

```
OBCU_ATO_Telegram_Builder.forwardingCommand,

step9:
```

**at** 8

**have**

```
DoorCommand.DT_DoorCommand_Close[
    OBCU_ITF_FWD_Platform_Door.
    WCU_OBCU_PlatformDoor_Out]
```

**from** [

```
step8 with [

(OBCU_ITF_FWD_Platform_Door.WCU_OBCU_PlatformDoor,
    OBCU_ATO_Telegram_Builder.OBCU_WCU_PlatformDoor)

]

]
```

**using**

```
OBCU_ITF_FWD_Platform_Door.forwardingCommand,
```

```
step10:
```

**at** 7

**have**

```
PSD_Authorization.DT_PSD_Authorization_Not_Authorize[
   OBCU_ITF_FWD_Position_Report.
   WCU_OBCU_PositionReport_Out]
```

**from** [

```
step4 with [

(OBCU_ITF_FWD_Position_Report.WCU_OBCU_PositionReport
   , OBCU_ATP_Telegram_Builder.
   OBCU_WCU_PositionReport)

]

]
```

**using**

```
OBCU_ITF_FWD_Position_Report.forwardAuthorization,

step11:
```

**at** 8

**have**

```
PSD_Authorization.DT_PSD_Authorization_Not_Authorize[
   Wayside_OPR_Telegram_Extractor.
   doorOpenAuthorization]
```

**from** [

```
step10 with [

(Wayside_OPR_Telegram_Extractor.
   OBCU_WCU_PositionReport,
   OBCU_ITF_FWD_Position_Report.
   WCU_OBCU_PositionReport_Out)
```

```
    ]
  ]
  using
  Wayside_OPR_Telegram_Extractor.forwardAuthorization,
step12:
  at 9
  have
  DoorCommand.DT_DoorCommand_Close[
      Wayside_OPD_Telegram_Extractor.psdOpenCommand]
  from [
    step9 with [
      (Wayside_OPD_Telegram_Extractor.OBCU_WCU_PlatformDoor
         , OBCU_ITF_FWD_Platform_Door.
         WCU_OBCU_PlatformDoor_Out)
    ]
  ]
  using
  Wayside_OPD_Telegram_Extractor.forwardCommand,
step13:
  at 11
  have
  Bool.false[Wayside_PSD_Control_Function.psdOpen]
  from [
    step12 with [
```

```
      (Wayside_PSD_Control_Function.psdOpenCommand,
         Wayside_OPD_Telegram_Extractor.psdOpenCommand)

   ],

   step11 with [

   (Wayside_PSD_Control_Function.doorOpenAuthorization,
      Wayside_OPR_Telegram_Extractor.
      doorOpenAuthorization)

   ]

   ]
```

**using**

```
Wayside_PSD_Control_Function.doorClosed,
```

```
step14:
```

**at** 13

**have**

```
Bool.true[Wayside_Out_Converter.PSD_O_PSD_Close] /\
   Bool.false[Wayside_Out_Converter.PSD_O_PSD_Open]
```

**from** [

```
step13 with [

(Wayside_Out_Converter.psdOpen,
   Wayside_PSD_Control_Function.psdOpen)

]

]
```

**using**

```
Wayside_Out_Converter.psdClose

}
```

```
    },

    // 2) If the train is at standstill and the position of
    //     the train doors match the position of the PSDs,
    //    then, the PSDs are opened.

    Contract PSDAreOpenIfNotMovingAndMatchingPosition {

     triggers {

      standstill: MovingStatus.
          DT_TrainMovingStatus_StandStill[
          OBCU_ATP_Door_Release_Controller.movingStatus],

      hasPSD: PlatformPSDStatus.
          DT_PlatformPSDStatus_Has_PSD[
          OBCU_ATP_Door_Release_Controller.platformPSDStatus
          ] at 1,

      inStopingPosition: Bool.true[Comparison.ls[Int.sub[
          OBCU_ATP_Door_Release_Controller.stoppingPoint,
          OBCU_ATP_Door_Release_Controller.
          stoppingWindowMinus],
          OBCU_ATP_Door_Release_Controller.trainPosition]]
          /\

      Bool.true[Comparison.gt[Int.sub[
          OBCU_ATP_Door_Release_Controller.stoppingPoint,
          OBCU_ATP_Door_Release_Controller.stopingWindowPlus
          ], OBCU_ATP_Door_Release_Controller.trainPosition
          ]] at 2

     }

     guarantees {

      Bool.true[Wayside_Out_Converter.PSD_O_PSD_Open] /\
          Bool.false[Wayside_Out_Converter.PSD_O_PSD_Close]

     }

     duration 13

     proof {
```

```
step1:

at 3

have

DoorReleaseCommand.DT_DoorReleaseCommand_Deactivate[
   OBCU_ATP_Door_Release_Controller.
   doorReleaseCommand] /\

DoorReleaseStatus.DT_DoorReleaseStatus_Locked[
   OBCU_ATP_Door_Release_Controller.doorReleaseStatus
   ]

from [

standstill, hasPSD, inStopingPosition

]

using

OBCU_ATP_Door_Release_Controller.
   doorsReleasedIfStandstillAndInStoppingWindow,

step2:

at 4

have

PSD_Authorization.DT_PSD_Authorization_Authorize[
   OBCU_ATP_PSD_Authorization_Controller.
   authorizePSDOpening]

from [

step1 with [

(OBCU_ATP_PSD_Authorization_Controller.
   doorReleaseStatus,
   OBCU_ATP_Door_Release_Controller.doorReleaseStatus
   )
```

```
      ]

    ]

  using

  OBCU_ATP_PSD_Authorization_Controller.notAuthorized,

  step3:

  at 5

  have

  PSD_Authorization.DT_PSD_Authorization_Authorize[
     OBCU_ATP_PSD_Authorization_State_Tracker.
     psdAuthorizationState]

  from [

    step2 with [

      (OBCU_ATP_PSD_Authorization_State_Tracker.
         authorizePSDOpening,
         OBCU_ATP_PSD_Authorization_Controller.
         authorizePSDOpening)

    ]

  ]

  using

  OBCU_ATP_PSD_Authorization_State_Tracker.
     forwardAuthorization,

  step4:

  at 6

  have

  PSD_Authorization.DT_PSD_Authorization_Authorize[
     OBCU_ATP_Telegram_Builder.OBCU_WCU_PositionReport]
```

```
        from [

        step3 with [

        (OBCU_ATP_Telegram_Builder.psdAuthorizationState,
            OBCU_ATP_PSD_Authorization_State_Tracker.
            psdAuthorizationState)

        ]

        ]

        using

        OBCU_ATP_Telegram_Builder.forwardAuthorization,

        step5:

        at 4

        have

        DoorCommand.DT_DoorCommand_Open[
            OBCU_ATO_Door_Open_Controller.doorOpenCommand]

        from [

        step1 with [

        (OBCU_ATO_Door_Open_Controller.doorReleaseStatus,
            OBCU_ATP_Door_Release_Controller.doorReleaseStatus
            )

        ]

        ]

        using

        OBCU_ATO_Door_Open_Controller.doorsClosed,

        step6:

        at 5
```

**have**

DoorCommand.DT_DoorCommand_Open[
    OBCU_ATO_PSD_Door_Controller.psdDoorCommand]

**from** [

step5 with [

(OBCU_ATO_PSD_Door_Controller.psdDoorCommand,
    OBCU_ATO_Door_Open_Controller.doorReleaseStatus)

]

]

**using**

OBCU_ATO_PSD_Door_Controller.forwardDoorCommand,

step7:

**at** 6

**have**

DoorCommand.DT_DoorCommand_Open[
    OBCU_ATO_PSD_Door_Command_State.
    psdDoorCommandState]

**from** [

step6 with [

(OBCU_ATO_PSD_Door_Command_State.openDoorCommand,
    OBCU_ATO_PSD_Door_Controller.psdDoorCommand)

]

]

**using**

OBCU_ATO_PSD_Door_Command_State.forwardingCommand,

```
step8:
```

**at** 7

**have**

```
DoorCommand.DT_DoorCommand_Open[
    OBCU_ATO_Telegram_Builder.OBCU_WCU_PlatformDoor]
```

**from** [

```
step7 with [

(OBCU_ATO_Telegram_Builder.psdDoorCommandState,
    OBCU_ATO_PSD_Door_Command_State.
    psdDoorCommandState)

]

]
```

**using**

```
OBCU_ATO_Telegram_Builder.forwardingCommand,

step9:
```

**at** 8

**have**

```
DoorCommand.DT_DoorCommand_Open[
    OBCU_ITF_FWD_Platform_Door.
    WCU_OBCU_PlatformDoor_Out]
```

**from** [

```
step8 with [

(OBCU_ITF_FWD_Platform_Door.WCU_OBCU_PlatformDoor,
    OBCU_ATO_Telegram_Builder.OBCU_WCU_PlatformDoor)

]

]
```

**using**

OBCU_ITF_FWD_Platform_Door.forwardingCommand,

step10:

**at** 7

**have**

PSD_Authorization.DT_PSD_Authorization_Authorize[
    OBCU_ITF_FWD_Position_Report.
    WCU_OBCU_PositionReport_Out]

**from** [

step4 with [

(OBCU_ITF_FWD_Position_Report.WCU_OBCU_PositionReport
    , OBCU_ATP_Telegram_Builder.
    OBCU_WCU_PositionReport)

]

]

**using**

OBCU_ITF_FWD_Position_Report.forwardAuthorization,

step11:

**at** 8

**have**

PSD_Authorization.DT_PSD_Authorization_Authorize[
    Wayside_OPR_Telegram_Extractor.
    doorOpenAuthorization]

**from** [

step10 with [

```
    (Wayside_OPR_Telegram_Extractor.
       OBCU_WCU_PositionReport,
       OBCU_ITF_FWD_Position_Report.
       WCU_OBCU_PositionReport_Out)

  ]

]
```

**using**

`Wayside_OPR_Telegram_Extractor.forwardAuthorization,`

`step12:`

**at** 9

**have**

```
DoorCommand.DT_DoorCommand_Open[
   Wayside_OPD_Telegram_Extractor.psdOpenCommand]
```

**from** [

`step9 with [`

```
(Wayside_OPD_Telegram_Extractor.OBCU_WCU_PlatformDoor
   , OBCU_ITF_FWD_Platform_Door.
   WCU_OBCU_PlatformDoor_Out)
```

]

]

**using**

`Wayside_OPD_Telegram_Extractor.forwardCommand,`

`step13:`

**at** 11

**have**

`Bool.true[Wayside_PSD_Control_Function.psdOpen]`

```
from [

step12 with [

(Wayside_PSD_Control_Function.psdOpenCommand,
    Wayside_OPD_Telegram_Extractor.psdOpenCommand)

],

step11 with [

(Wayside_PSD_Control_Function.doorOpenAuthorization,
    Wayside_OPR_Telegram_Extractor.
    doorOpenAuthorization)

]

]

using

Wayside_PSD_Control_Function.doorClosed,

step14:

at 13

have

Bool.true[Wayside_Out_Converter.PSD_O_PSD_Close] /\
    Bool.false[Wayside_Out_Converter.PSD_O_PSD_Open]

from [

step13 with [

(Wayside_Out_Converter.psdOpen,
    Wayside_PSD_Control_Function.psdOpen)

]

]

using
```

```
    Wayside_Out_Converter.psdClose

  }

},

// 3) If the train doors open on the right hand side,
    the platform must be on the right hand

Contract trainOpensTheDoorOnTheRightSide {

 triggers {

  openRightDoorStrategy: DoorOpenStrategy.
     DT_DoorOpenStrategy_Right[
     OBCU_ATP_Door_Release_Controller.platformSide]

 }

 guarantees {

  DoorReleaseCommand.DT_DoorReleaseCommand_Right[
     OBCU_ATP_Door_Release_Controller.
     doorReleaseCommand]

 }

 duration 1

 proof {

  step1:

  at 1

  have

  DoorReleaseCommand.DT_DoorReleaseCommand_Right[
     OBCU_ATP_Door_Release_Controller.
     doorReleaseCommand]

  from [
```

```
     openRightDoorStrategy

    ]

   using

    OBCU_ATP_Door_Release_Controller.openingRightDoorSide

  }

 },

 // 4) If permissive release and the train is at
    standstill then both doors are open

 Contract trainOpensTheDoorOnTheRightSide {

  triggers {

   standstill: MovingStatus.
      DT_TrainMovingStatus_StandStill[
      OBCU_ATP_Door_Release_Controller.movingStatus],

   permissiveButtonPresset: Bool.true[
      OBCU_ATP_Door_Release_Controller.
      cabPremissiveDoorButton]

  }

  guarantees {

   Bool.true[Wayside_Out_Converter.PSD_O_PSD_Open] /\
      Bool.false[Wayside_Out_Converter.PSD_O_PSD_Close]
      /\

   DoorReleaseCommand.DT_DoorReleaseCommand_Right[
      OBCU_ATP_Door_Release_Controller.
      doorReleaseCommand]

  }

  duration 1

  proof {
```

```
     step1:

     at 3

     have

     DoorReleaseCommand.DT_DoorReleaseCommand_Both[
         OBCU_ATP_Door_Release_Controller.
         doorReleaseCommand] /\

     DoorReleaseStatus.DT_DoorReleaseStatus_Released[
         OBCU_ATP_Door_Release_Controller.doorStatus] /\

     Bool.true[OBCU_ATP_Door_Release_Controller.
         releaseForcedByPermissiveDoorButton]

     from [

     standstill, permissiveButtonPresset

     ]

     using

     OBCU_ATP_Door_Release_Controller.
         doorsReleasedIfStandstillAndInStoppingWindow

  }

 },

// 5) When the train indicates that the doors are
    closed, PSDs are closed

 Contract PSDAreClosedWhenTrainGivesClosedIndication {

  triggers {

   trigger1: DoorStatus.DT_DoorStatus_DoorsVitalClosed[
       OBCU_ATP_In_Converter.TCL_I_Door_Closed_Indication
       ]

  }
```

```
guarantees {

 Bool.true[Wayside_Out_Converter.PSD_O_PSD_Close] /\
     Bool.false[Wayside_Out_Converter.PSD_O_PSD_Open]

}

duration 14

proof {

 step1:

 at 1

 have

 DoorStatus.DT_DoorStatus_DoorsVitalClosed[
     OBCU_ATP_In_Converter.doorStatus]

 from [

 trigger1

 ]

 using

 OBCU_ATP_In_Converter.forwardDoorStatus,

 step2:

 at 4

 have

 DoorReleaseCommand.DT_DoorReleaseCommand_Deactivate[
     OBCU_ATP_Door_Release_Controller.
     doorReleaseCommand] /\

 DoorReleaseStatus.DT_DoorReleaseStatus_Locked[
     OBCU_ATP_Door_Release_Controller.doorReleaseStatus
     ]

 from [
```

```
                    step1 with [

(OBCU_ATP_Door_Release_Controller.doorStatus,
    OBCU_ATP_In_Converter.doorStatus)

]

]
```

**using**

```
OBCU_ATP_Door_Release_Controller.
    doorsClosedWhenClosedIndication,

step3:
```

**at** 5

**have**

```
PSD_Authorization.DT_PSD_Authorization_Not_Authorize[
    OBCU_ATP_PSD_Authorization_Controller.
    authorizePSDOpening]
```

**from** [

```
step2 with [

(OBCU_ATP_PSD_Authorization_Controller.
    doorReleaseStatus,
    OBCU_ATP_Door_Release_Controller.doorReleaseStatus
    )

]

]
```

**using**

```
OBCU_ATP_PSD_Authorization_Controller.notAuthorized,

step4:
```

**at** 6

**have**

PSD_Authorization.DT_PSD_Authorization_Not_Authorize[
    OBCU_ATP_PSD_Authorization_State_Tracker.
    psdAuthorizationState]

**from** [

step3 with [

(OBCU_ATP_PSD_Authorization_State_Tracker.
    authorizePSDOpening,
    OBCU_ATP_PSD_Authorization_Controller.
    authorizePSDOpening)

]

]

**using**

OBCU_ATP_PSD_Authorization_State_Tracker.
    forwardAuthorization,

step5:

**at** 7

**have**

PSD_Authorization.DT_PSD_Authorization_Not_Authorize[
    OBCU_ATP_Telegram_Builder.OBCU_WCU_PositionReport]

**from** [

step4 with [

(OBCU_ATP_Telegram_Builder.psdAuthorizationState,
    OBCU_ATP_PSD_Authorization_State_Tracker.
    psdAuthorizationState)

]

]

```
using
OBCU_ATP_Telegram_Builder.forwardAuthorization,
step6:
at 5
have
DoorCommand.DT_DoorCommand_Close[
    OBCU_ATO_Door_Open_Controller.doorOpenCommand]
from [
step2 with [
(OBCU_ATO_Door_Open_Controller.doorReleaseStatus,
    OBCU_ATP_Door_Release_Controller.doorReleaseStatus
    )
]
]
using
OBCU_ATO_Door_Open_Controller.doorsClosed,
step7:
at 6
have
DoorCommand.DT_DoorCommand_Close[
    OBCU_ATO_PSD_Door_Controller.psdDoorCommand]
from [
step6 with [
(OBCU_ATO_PSD_Door_Controller.psdDoorCommand,
    OBCU_ATO_Door_Open_Controller.doorReleaseStatus)
```

```
        ]

    ]

    using
    OBCU_ATO_PSD_Door_Controller.forwardDoorCommand,

step8:

    at 7

    have
    DoorCommand.DT_DoorCommand_Close[
        OBCU_ATO_PSD_Door_Command_State.
        psdDoorCommandState]

    from [

        step7 with [

            (OBCU_ATO_PSD_Door_Command_State.openDoorCommand,
                OBCU_ATO_PSD_Door_Controller.psdDoorCommand)

        ]

    ]

    using
    OBCU_ATO_PSD_Door_Command_State.forwardingCommand,

step9:

    at 8

    have
    DoorCommand.DT_DoorCommand_Close[
        OBCU_ATO_Telegram_Builder.OBCU_WCU_PlatformDoor]

    from [
```

```
      step8 with [

(OBCU_ATO_Telegram_Builder.psdDoorCommandState,
    OBCU_ATO_PSD_Door_Command_State.
    psdDoorCommandState)

]

]
```

**using**

```
OBCU_ATO_Telegram_Builder.forwardingCommand,

step10:
```

**at** 9

**have**

```
DoorCommand.DT_DoorCommand_Close[
    OBCU_ITF_FWD_Platform_Door.
    WCU_OBCU_PlatformDoor_Out]
```

**from** [

```
step9 with [

(OBCU_ITF_FWD_Platform_Door.WCU_OBCU_PlatformDoor,
    OBCU_ATO_Telegram_Builder.OBCU_WCU_PlatformDoor)

]

]
```

**using**

```
OBCU_ITF_FWD_Platform_Door.forwardingCommand,

step11:
```

**at** 8

**have**

```
PSD_Authorization.DT_PSD_Authorization_Not_Authorize[
   OBCU_ITF_FWD_Position_Report.
   WCU_OBCU_PositionReport_Out]
```

**from** [

step5 with [

(OBCU_ITF_FWD_Position_Report.WCU_OBCU_PositionReport
   , OBCU_ATP_Telegram_Builder.
   OBCU_WCU_PositionReport)

]

]

**using**

OBCU_ITF_FWD_Position_Report.forwardAuthorization,

step12:

**at** 9

**have**

PSD_Authorization.DT_PSD_Authorization_Not_Authorize[
   Wayside_OPR_Telegram_Extractor.
   doorOpenAuthorization]

**from** [

step11 with [

(Wayside_OPR_Telegram_Extractor.
   OBCU_WCU_PositionReport,
   OBCU_ITF_FWD_Position_Report.
   WCU_OBCU_PositionReport_Out)

]

]

**using**

```
Wayside_OPR_Telegram_Extractor.forwardAuthorization,

step13:

at 10

have

DoorCommand.DT_DoorCommand_Close[
   Wayside_OPD_Telegram_Extractor.psdOpenCommand]

from [

step10 with [

(Wayside_OPD_Telegram_Extractor.OBCU_WCU_PlatformDoor
   , OBCU_ITF_FWD_Platform_Door.
   WCU_OBCU_PlatformDoor_Out)

]

]

using

Wayside_OPD_Telegram_Extractor.forwardCommand,

step14:

at 12

have

Bool.false[Wayside_PSD_Control_Function.psdOpen]

from [

step12 with [

(Wayside_PSD_Control_Function.doorOpenAuthorization,
   Wayside_OPR_Telegram_Extractor.
   doorOpenAuthorization)

],
```

```
        step13 with [

        (Wayside_PSD_Control_Function.psdOpenCommand,
            Wayside_OPD_Telegram_Extractor.psdOpenCommand)

        ]

      ]

      using

      Wayside_PSD_Control_Function.doorClosed,

      step15:

      at 14

      have

      Bool.true[Wayside_Out_Converter.PSD_O_PSD_Close] /\
          Bool.false[Wayside_Out_Converter.PSD_O_PSD_Open]

      from [

        step14 with [

        (Wayside_Out_Converter.psdOpen,
            Wayside_PSD_Control_Function.psdOpen)

        ]

      ]

      using

      Wayside_Out_Converter.psdClose
    }
  }
 }
}
```